# Finite Volume Adaptive Solutions using SIMPLE as Smoother

**Alexandros Syrakos, Apostolos Goulas**
Department of Mechanical Engineering, Aristotle University of Thessaloniki, Macedonia, Greece.

**ABSTRACT**

This paper describes a new multilevel procedure which can solve the discrete Navier-Stokes system arising from finite volume discretisations on composite grids which may consist of more than one level. SIMPLE is used and tested as the smoother, but the multilevel procedure is such that it does not exclude the use of other smoothers. Local refinement is guided by a criterion based on an estimate of the truncation error. The numerical experiments presented test not only the behaviour of the multilevel algebraic solver but also the efficiency of local refinement based on this particular criterion.

## 1. INTRODUCTION

The SIMPLE algorithm ([1]) and its variants (see [2] for descriptions and references) are popular methods for the solution of discrete non-linear systems of equations arising from the discretisation of the incompressible Navier – Stokes equations. Like most single-grid solution methods they have the disadvantage that the larger the grid (in terms of number of nodes), the larger the number of iterations required to attain a specified level of convergence.

Multigrid ([3], [4], [5]) is a class of methods which overcome this disadvantage by using a series of progressively coarser grids in addition to the finest one. Any solver for systems of algebraic equations may be used in the context of a multigrid method, and in this case it is called a *smoother* rather than a solver. In a multigrid method, the error of a current approximation compared to the exact solution of the system is considered to be composed of Fourier components, and each grid is responsible for the reduction of those components which may be represented on it and which oscillate rapidly compared to the grid spacing. The efficiency of the method depends on the ability of the iterative solver to reduce the rapidly oscillating components of the error faster than the smooth components on each grid. After a number of iterations of such a solver the error appears smoother than before, and therefore in the context of multigrid such solvers are called smoothers. By transferring the solution procedure between grids of various densities, every component of the error will appear oscillatory on some grid, where it will be efficiently reduced by the smoother. Therefore, ideally, the number of iterations (or *cycles*) required in multigrid methods to attain a specified level of convergence is small and independent of the density of the finest grid.

Originally Gauss-Seidel and line relaxation smoothers were used, but beginning in the late 1980s several studies have shown that SIMPLE-like procedures also have smoothing properties. In [6] and [7] this was shown for staggered grids, while in [8] an algorithm was derived for colocated grids. In [9] the method was tried on curvilinear grids, demonstrating high efficiency, and in [10] and [11] the method was tried on three-dimensional cases. In [12], [13] and [14] it was shown that the method can handle turbulence models successfully, although the gains are not in general as high as in the laminar case. In [15] the effect of the discretisation scheme for the convection terms on the overall efficiency was studied. In [16] and [17] it was shown that the performance of SIMPLE as a smoother is comparable to that of SCGS (Gauss-Seidel – like smoother) and SCAL (line-relaxation – type smoother). Also, in [18] SIMPLE is compared with other methods of the same family and it is shown that their performance as smoothers is not significantly different. A more specialised use of the procedure is presented in [19], where SIMPLE is the smoother of a multigrid procedure which is used as a preconditioner for a Newton-Krylov method.

In all these studies, except maybe [13], each of the grids used by the multigrid procedure covers the entire computational domain. There are cases though where it would be convenient that each grid is allowed to cover a subset of the domain. Such grids are more appropriately termed *levels*, and the corresponding solution techniques are called *multilevel* rather than multigrid techniques. Multilevel techniques are convenient in the case of locally refined grids, where finer levels are added on top of an original grid only in regions of the flow field where enhanced spatial resolution is required. Multilevel techniques date back to the work of Brandt [4], who defined the technique known as MLAT (MultiLevel Adaptive Technique) for finite difference discretisations. In the case of finite volume discretisations some caution is needed in constructing the multigrid equations for the control volumes which are located at interfaces between different levels. In [5] a method is proposed for composite grids (i.e. grids composed of different levels) whose nodes identify with the centres of the control volumes. In the present work a method will be proposed for grids whose nodes identify with the vertices of the control volumes.

The classical SIMPLE/multigrid formulation as described in the studies mentioned above cannot readily be applied in the case of locally refined (composite) grids. In [13] results of using SIMPLE as smoother on composite grids are presented, but the method described is precisely the classic one, and the difficulties which arise at the interfaces between different levels are not addressed. These will be addressed in the present work. A different approach is adopted in [20], where the composite grid is partitioned into blocks, each assigned to a different processor, and each block is covered by a single level. The classic SIMPLE/multigrid method is used within each block, and information is exchanged at the interfaces between blocks, between successive multigrid cycles. However, in the present work the composite grid will not be partitioned into blocks but treated as a whole.

In this paper the emphasis is on how to extend the multigrid procedure to be used on composite grids, but one also has to choose an appropriate refinement criterion on which the construction of the composite grid will be based. In some studies, e.g. [20], [21], the grid is refined before the calculations are performed, at regions where the flow field is expected to exhibit strong variation e.g. boundary layers. A more useful approach is to solve the equations on a given grid and then refine those regions of the grid which are indicated by a criterion which assesses the quality of the solution. This can be done by examining the variation of the solution in relation with the grid spacing. For example, a second order accurate discretisation, like the one used in this paper, is constructed by assuming that the control volumes are small enough such that the flow variables vary almost linearly within the neighbourhood of each control volume. Therefore, after a solution has been obtained on a given grid, the refinement criterion can refine those control volumes where this assumption does not hold within a selected tolerance. Such refinement criteria can be based on estimates of the truncation error (e.g. [22], [5]), the finite element residual (e.g. [23], [24]) or some other similar quantity (e.g. [13]). These quantities have the same dimensions as the flux of the transported quantity (e.g. mass or momentum) but some criteria (e.g. [24], [13]) normalise them by some factor (usually the main diagonal coefficient of the linearised algebraic system) to convert them into a quantity which has the same dimensions as the flow variables. In this paper the refinement criterion is based on the truncation error estimate presented in [25], which is not normalised but the goal is to reduce the integral of the truncation error in the domain below some selected value.

## 2. DISCRETISATION OF THE DOMAIN INTO FINITE VOLUMES

The continuous space of the flow field under study is divided into quadrilateral control volumes (CVs), which are organised into levels (see Figure 1). The grid is constructed as follows: First, level 1 is constructed (say, by a structured grid generator) which covers the entire computational domain and is relatively coarse. Then a second level is added, whose CVs are formed by subdivision of some (or all) of the CVs of level 1. The subdivision procedure produces four (4) "children" CVs out of each chosen CV of the previous level, by joining the parent CV centre with the midpoints of its 4 sides. Level 3 is similarly constructed by subdividing some (or all) of the CVs of level 2 and so on. A restriction is posed on this procedure: at an interface between two different levels, say $k$ and $l$, the two levels which meet must be successive, i.e. $|k–l|=1$.

Once a CV is subdivided into children it is not destroyed but it is retained in the data structure because it will be needed by the multilevel procedure. Therefore, each level consists of its *global* part, i.e. the CVs which do not have children, and its *local* part, i.e. the CVs which have children. The sum of the global parts of all levels forms the *composite grid*, which is the actual grid used for the discretisation of the Navier-Stokes equations.

A composite grid will be denoted by a letter such as $h$ which will also be abstractly interpreted as the distribution of the sizes of the CVs of the given composite grid in the computational domain. A multiple of $h$, say $a \cdot h$, will denote a composite grid whose CV at any point of the domain has size (length) $a$ times the size of, and is geometrically similar to, the corresponding CV of grid $h$ at the same location. Further, it is required that the CV relates to its neighbours in a similar way that the corresponding CV of grid $h$ relates to its own neighbours. For an arbitrary grid $h$ it is generally not possible to find a grid which fulfils the requirements for being $a \cdot h$, but often one may find grids which approximately fulfil these requirements. An important special case is the following:

Some composite grids have an *underlying grid* which is formed by removing all global CVs from the original grid. Therefore, the global CVs of the underlying grid are precisely those local CVs of the original grid which do not have "grandchildren". For an underlying grid to exist, all local CVs of the original grid must have only local siblings (siblings are the CVs which have a common parent). For example, the grid of Figure 1 does not have an underlying grid. This grid is reproduced in Figure 2(a), where the global siblings of local CVs are highlighted in grey. Once these CVs are refined, as in Figure 2(b), the underlying grid shown in Figure 2(c) becomes available. One can ensure that an underlying grid exists at all stages of the finite volume solution by requiring that whenever a CV is marked for refinement, say due to high truncation error, all it's siblings are marked as well. In the present method, the underlying grid is used to estimate the truncation error.

In the present method the CVs are considered to be logically polygonal, an approach suggested for example in [13], [2] and [26]. According to this approach the surface which forms the boundary of each CV is considered to consist of a number of *faces*, each of which separates the given CV from another single distinct CV. Therefore, although the CVs of the present method are quadrilaterals, those CVs which are located at interfaces between different levels may have more than 4 faces. For example, CV $P$ of Figure 3 has 6 faces and 6 neighbours. The faces which separate two CVs of different level will be called *exterior*, while the rest will be called *interior*, except for *boundary* faces which coincide with the domain boundary.

If $\phi$ is a function defined on the computational domain and the computational domain is discretised by a grid $h$, then the *grid function* $\phi_h$ is the discrete function which assumes the values of $\phi$ at the centres of the CVs of grid $h$. The $P$-th component of $\phi_h$, which equals the value of $\phi$ at the centre of CV $P$, is denoted as $\phi_{h,P}$ or $(\phi_h)_P$.

## 3. THE EQUATIONS AND THEIR DISCRETISATION

In Cartesian coordinates, the two-dimensional stationary incompressible Navier – Stokes equations and the continuity equation integrated over a control volume $P$ may be written as:

$$N_P^x(u,v,p) \equiv \sum_{f \in f_P} M_f^x = 0 \quad , \quad N_P^y(u,v,p) \equiv \sum_{f \in f_P} M_f^y = 0 \quad , \quad N_P^c(u,v) \equiv \sum_{f \in f_P} F_f = 0 \quad (3.1)$$

where $u$ and $v$ are the $x$– and $y$– components of the velocity vector $V = u\mathbf{i} + v\mathbf{j}$, $p$ is the pressure, $f_P$ is the set of all faces of CV $P$, and $F_f$, $M_f^x$, $M_f^y$ are respectively the mass flux and net momentum fluxes minus forces in the $x$–, $y$– directions through face $f$ defined as:

$$M_f^x(u,v,p) \equiv \int_{S_f} \rho V \cdot \mathbf{n}\, u\, dS \;-\; \int_{S_f} \mu\left(\nabla u + \frac{\partial u}{\partial x}\mathbf{i} + \frac{\partial v}{\partial x}\mathbf{j}\right)\cdot \mathbf{n}\, dS \;+\; \int_{S_f} p\mathbf{i}\cdot \mathbf{n}\, dS \qquad (3.2)$$

$$M_f^y(u,v,p) \equiv \int_{S_f} \rho V \cdot \mathbf{n}\, v\, dS \;-\; \int_{S_f} \mu\left(\nabla v + \frac{\partial u}{\partial y}\mathbf{i} + \frac{\partial v}{\partial y}\mathbf{j}\right)\cdot \mathbf{n}\, dS \;+\; \int_{S_f} p\mathbf{j}\cdot \mathbf{n}\, dS \qquad (3.3)$$

$$F_f(u,v) \equiv \int_{S_f} \rho V \cdot \mathbf{n}\, dS \qquad (3.4)$$

where $S_f$ is the surface of face $f$ and $\mathbf{n}$ is the normal unit vector at each point of the surface, $\mathbf{i}$ and $\mathbf{j}$ are the unit vectors in the $x$– and $y$– directions respectively, $\rho$ is the density and $\mu$ is the viscosity of the fluid. The normal unit vector $\mathbf{n}$ is assumed to be directed outward of the CV under consideration, and therefore if $f$ is not a boundary face then (3.2) – (3.4) are defined with opposite sign for the CV which lies on the other side of $f$.

Equations (3.1) are discretised as follows: First of all, the spatial derivatives of the variables $(u, v, p)$ at each CV centre are approximated from their values at the centre of the CV and at the centres of its neighbours, using the least-squares method described in [13] and in [26]. This results in a discrete gradient operator $\nabla_h$, with its two Cartesian components $\nabla_h^x, \nabla_h^y$. For more details on how to construct this operator and about its accuracy see [27].

Then, the fluxes and forces on each face are discretised: Suppose a face $f$ separates CVs $P$ and $N$ with centres $\mathbf{P}$ and $\mathbf{N}$ respectively, with the surface normal pointing from $P$ to $N$ (see Figure 4). The centre of the face is denoted by $\mathbf{c}$, and $\mathbf{c'}$ denotes the point on the line $\mathbf{PN}$ which is closest to $\mathbf{c}$. Also points $\mathbf{P'}$ and $\mathbf{N'}$ are such that the segment $\mathbf{P'N'}$ has the same length as $\mathbf{PN}$ and it is perpendicular to the face $f$ and its midpoint is $\mathbf{c}$. Let an overbar denote a value obtained at point $\mathbf{c'}$ by linear interpolation from the values at points $\mathbf{P}$ and $\mathbf{N}$, and define approximate values of a function $\phi$ at points $\mathbf{c}$ and $\mathbf{P'}$ as:

$$\phi_{h,c} \equiv \overline{\phi_{h,c'}} + \overline{(\nabla_h \phi_h)}_{c'} \cdot (\mathbf{c}-\mathbf{c'}) \qquad (3.5)$$

$$\phi_{h,P'} \equiv \phi_{h,P} + (\nabla_h \phi_h)_P \cdot (\mathbf{P'}-\mathbf{P}) \qquad (3.6)$$

Then the fluxes $M_f^x$ (3.2) and $F_f$ (3.4) are approximated by:

$$M_{h,f}^x(u_h,v_h,p_h) \equiv F_{h,f}\cdot u_{h,c} \;-\; \mu S_f \frac{u_{h,N'}-u_{h,P'}}{|\mathbf{N'}-\mathbf{P'}|} \;-\; \mu S_f\left[\overline{(\nabla_h^x u_h)}_{c'} n_f^x + \overline{(\nabla_h^x v_h)}_{c'} n_f^y\right] \\ + p_{h,c} n_f^x S_f \qquad (3.7)$$

$$F_{h,f}(u_h,v_h,p_h) \equiv \rho S_f \cdot \left(V_{h,c}\cdot \mathbf{n}_f \;+\; a_{mi}\frac{S_f}{A_f}\left[(p_{h,P}-p_{h,N}) - (\nabla_h p_h)_f \cdot (\mathbf{P}-\mathbf{N})\right]\right) \qquad (3.8)$$

where:

$$(\nabla_h p_h)_f \equiv \frac{1}{2}\left[(\nabla_h p_h)_P + (\nabla_h p_h)_N\right] \qquad (3.9)$$

$$A_f \equiv \rho S_f |V_{h,c}\cdot \mathbf{n}_f| \;+\; \rho S_V |V_{h,c}\cdot \mathbf{rot}(\mathbf{n}_f)| \;+\; 2\mu\cdot\left[\frac{S_f}{S_V} + \frac{S_V}{S_f}\right] \qquad (3.10)$$

In the above, $n_f^x$, $n_f^y$ are the Cartesian components of the unit vector $\mathbf{n}_f$ which is perpendicular to $f$ and points from $P$ to $N$. Also, $\mathbf{V}_{h,c} = u_{h,c}\cdot\mathbf{i} + v_{h,c}\cdot\mathbf{j}$, $S_V = (\mathbf{N}-\mathbf{P})\cdot\mathbf{n}_f$, and **rot**: $\mathbb{R}^2 \to \mathbb{R}^2$ is a function which rotates a vector by 90°. The flux $M_f^y$ (3.3) is approximated by a flux $M_{h,f}^y$ which is defined similarly to (3.7). The definition (3.8) of the discrete mass flux is taken from [25], and it contains a pressure term to avoid the appearance of pressure oscillations in the discrete solution. The real factor $a_{mi}$ is included for better control of the pressure term. Usually one can use $a_{mi}=1$, except for some rare cases of very coarse grids where $a_{mi} < 1$ is necessary to avoid convergence problems (see [25] or [27] for more details). In the experiments of Section 6, $a_{mi}=1$ is used unless otherwise specified.

With the above definitions, the exact operators $N^x$, $N^y$, $N^c$ (3.1) are approximated by the algebraic operators $N_h^x$, $N_h^y$, $N_h^c$, which differ by truncation errors $\tau_h^x$, $\tau_h^y$, $\tau_h^c$ respectively:

$$N_{h,P}^x(u_h,v_h,p_h) \equiv \sum_{f\in f_P} M_{h,f}^x \quad , \quad N_{h,P}^c(u_h,v_h,p_h) \equiv \sum_{f\in f_P} F_{h,f} \qquad (3.11)$$

$$\begin{aligned} N_{h,P}^x(u_h,v_h,p_h) + \tau_{h,P}^x(u,v,p)\cdot\Delta\Omega_P &= N_P^x(u,v,p) \\ N_{h,P}^c(u_h,v_h,p_h) + \tau_{h,P}^c(u,v,p)\cdot\Delta\Omega_P &= N_P^c(u,v) \end{aligned} \qquad (3.12)$$

where $\Delta\Omega_P$ is the volume of CV $P$, and similarly for $N^y$, $N_h^y$, $\tau_h^y$. If (3.12) were substituted into (3.1) and the system was solved one would obtain the exact values $u_h$, $v_h$, $p_h$ at the CV centres. However, this is not possible since the truncation errors are not known. According to the finite volume methodology, the truncation errors are dropped from (3.12) under the assumption that they have small magnitude, and instead of (3.1) the following algebraic system is solved:

$$N_h^x(u_h^*,v_h^*,p_h^*)=0 \quad , \quad N_h^y(u_h^*,v_h^*,p_h^*)=0 \quad , \quad N_h^c(u_h^*,v_h^*,p_h^*)=0 \qquad (3.13)$$

The solution $u_h^*$, $v_h^*$, $p_h^*$ of (3.13) differs from the exact solution $u_h$, $v_h$, $p_h$ of (3.1) by the discretisation error $\varepsilon_h^u = u_h - u_h^*$, $\varepsilon_h^v = v_h - v_h^*$, $\varepsilon_h^p = p_h - p_h^*$.

For arbitrary grids, it is shown in [27] that the contributions of the convection and pressure terms of (3.7) to the truncation error have a magnitude of $O(h)$, while the contributions of the viscous terms have a magnitude of $O(1)$. However on structured grids which come from the discretisation of smooth fields of dimensionless curvilinear coordinates, where the grid skewness and expansion vary smoothly between opposite faces of a CV and tend to zero with grid refinement, parts of these contributions cancel out between opposite faces of a CV and what remains is that $\tau_{h,P}^x$, $\tau_{h,P}^y \in O(h^2)$ for each CV $P$. (The skewness of a face may be defined as the ratio $|c-c'|/|N-P|$, and the expansion as the ratio $|(N+P)/2 - c'|/|N-P|$ – see Figure 4). Similarly, in the general case $\tau_h^c \in O(h)$ but on smooth structured grids $\tau_h^c \in O(h^2)$. Therefore many numerical experiments, e.g. in [25], [27], show that on smooth structured grids $\varepsilon_h^u$, $\varepsilon_h^v$, $\varepsilon_h^p \in O(h^2)$ also. A complication arises at boundary CVs, where $\tau_{h,P}^x$, $\tau_{h,P}^y \in O(1)$ and $\tau_{h,P}^c \in O(h)$ because (3.7) or (3.8) are not used for the boundary face, and therefore the truncation error contributions of (3.7) and (3.8) of the face which is opposite to the boundary face are not cancelled out. However, as numerical experiments have shown, it appears that these truncation errors of the boundary CVs do not affect the overall $O(h^2)$ convergence rate of the discretisation errors. Theoretical explanations for this are proposed in [28] and [29] for the case of the diffusion terms.

For a composite grid $h$, let $V_h^e$ be the set of all CVs which have an exterior face, and $V_h^{en}$ be the set of all CVs which do not belong to $V_h^e$ but have at least one immediate neighbour which belongs to $V_h^e$. Then it is shown in [27] that $\tau_{h,P}^x$, $\tau_{h,P}^y \in O(1)$ for all $P \in V_h^e \cup V_h^{en}$ and $\tau_{h,P}^c \in O(h)$ for all $P \in V_h^e$. This is because skewness and expansion of exterior faces are large and do not reduce with grid refinement (see Figure 3). However, experiments in Section 6 of the present work suggest that again this does not affect the $O(h^2)$ convergence rate of the discretisation error. This may be due to cancellation between parts of the truncation errors of the coarse CV and the pair of fine CVs which meet at a level interface (e.g. CVs $P$, $N_3$, $N_4$ of Figure 3). Of course these truncation errors occur at different locations in space, but as the grid spacing $h$ reduces such triads of CVs tend to identify with single points in space.

## 4. TRUNCATION ERROR ESTIMATE AND LOCAL GRID REFINEMENT CRITERION

The refinement criterion which will be used in the context of this work is based on an estimate of the truncation error. The truncation error is estimated with the method proposed in [25]. The truncation error estimate $\tau_h^{x*}$ is given by:

$$\tau_h^{x*} = -\frac{1}{2^p - 1} I_{2h}^h \left[ \left[ \Delta \Omega \right]^{-1} \cdot N_{2h}^x \left( I_h^{2h} u_h^*, I_h^{2h} v_h^*, I_h^{2h} p_h^* \right) \right] \qquad (4.1)$$

The estimates $\tau_h^{y*}$, $\tau_h^{c*}$ are given by the same formula with $N_{2h}^x$ replaced by $N_{2h}^y$ and $N_{2h}^c$ respectively. In the above, grid $2h$ is the underlying grid of grid $h$ and the operators $N_{2h}^x$, $N_{2h}^y$, $N_{2h}^c$ are defined on grid $2h$ using the same discretisation schemes as $N_h^x$, $N_h^y$, $N_h^c$. $[\Delta\Omega]^{-1}$ is a diagonal matrix whose $P$-th diagonal element equals $1/\Delta\Omega_P$, the reciprocal of the volume of CV $P$ of grid $2h$. The linear operator $I_h^{2h}$ is called a restriction operator and it transfers a grid function from grid $h$ to grid $2h$. Likewise the prolongation operator $I_{2h}^h$ transfers a grid function from grid $2h$ to grid $h$. It was found that the particular choice of the restriction and prolongation operators does not have a strong impact on the overall efficiency of the local refinement method. The 3$^{rd}$ order accurate restriction operator and the prolongation operator used in [25] are also adopted in the present work. Finally, $p$ is the order of accuracy of the discretisation, which is assumed to be $p = 2$.

The local refinement criterion used here has as an ultimate goal to reduce the integral of the truncation error below some limit, while simultaneously distributing it roughly equally among all CVs of the grid. The limit can be set as follows: Given reference momentum and mass fluxes $Q_{mom}$ and $Q_{mas}$ respectively, which are of the order of the momentum and mass fluxes in the computational domain, and an appropriate real number $r_\tau \ll 1$, the goal is to reduce the integrals of the momentum and continuity truncation errors below $r_\tau \cdot Q_{mom}$ and $r_\tau \cdot Q_{mas}$ respectively. The (approximate) integrals of the truncation errors are defined as $\sum_P |\tau_{h,P}| \cdot \Delta\Omega_P$ for each truncation error, where summation is over all CVs $P$ of grid $h$, and $\Delta\Omega_P$ is the volume of CV $P$. This goal may be achieved by marking for refinement every CV $P$ of grid $h$ which does not fulfil the following condition:

$$\left| \tau_{h,P}^{x,y} \right| \cdot \Delta\Omega_P \leq \frac{r_\tau \cdot Q_{mom}}{\#V_h} \quad , \quad \left| \tau_{h,P}^c \right| \cdot \Delta\Omega_P \leq \frac{r_\tau \cdot Q_{mas}}{\#V_h} \qquad (4.2)$$

where $\#V_h$ is the number of CVs in grid $h$. When (4.2) is satisfied by all CVs of the grid then $\sum_P |\tau_{h,P}| \cdot \Delta\Omega_P \leq r_\tau \cdot Q$ as can be easily seen by summing (4.2) of all CVs of the grid ($\#V_h$ in number). Of course since the exact truncation errors are not known one has to use the estimates (4.1). After refinement of all CVs which do not satisfy (4.2) a new grid arises on which the equations are solved. Then again refinement takes place according to the criterion (4.2) and so on, until a grid is reached such that all its CVs satisfy (4.2). The criterion (4.2) is dynamic in the sense that a CV of a grid $h$ which satisfies (4.2) may not satisfy it in a subsequent grid $h'$, and thus be refined, because the number of CVs has increased ($\#V_{h'} >$

#$V_h$). In this way the procedure is aimed at reducing the integrals of the truncation errors below $r_\tau \cdot Q$ by refining the CVs with the largest contribution to these integrals.

The discrete operators (3.11) of any particular CV, and therefore also the truncation errors, depend not only on the size of $P$ itself but also on the sizes of its neighbours, so if $P$ does not satisfy (4.2) then its neighbours will be refined as well.

A complication arises near the interfaces between different levels of the composite grid where, as has been mentioned before, $\tau_h^x$, $\tau_h^y \in O(1)$ ($p = 0$) and $\tau_h^c \in O(h)$ ($p = 1$). Therefore, as far as the truncation error is concerned the assumption that $p=2$ in (4.1) is not accurate there. Furthermore in these regions the truncation error distribution is discontinuous and so $I_{2h}^h$ in (4.1), which is based on an assumption of smoothness, is not appropriate. Despite these problems, numerical experiments show that the estimate (4.1) predicts the zero-order accuracy of the discretisation, i.e. that the truncation error estimate is large and does not reduce with refinement. This poses a problem to the local refinement procedure: According to the criterion (4.2) the CVs of the set $V_h^e \cup V_h^{ne}$ must be refined, but this offers no real benefit as the truncation error will not reduce ($\tau_h \in O(1)$). On the next grid $h'$ the errors $\tau_{h'}^{x*}$, $\tau_{h'}^{y*}$ will still be high at the CVs of $V_{h'}^e \cup V_{h'}^{ne}$ causing refinement to take place and so on, resulting in perpetual refinement with no real benefit. This behaviour is indeed observed in practice. To overcome this problem in the present work it has been chosen not to allow refinement at these regions. Since the truncation error is effectively estimated on the underlying grid $2h$, refinement must not be allowed at the CVs of grid $h$ which cover $V_{2h}^e \cup V_{2h}^{ne}$. That is, refinement is not allowed within a depth of 4 CVs from the level interface.

A similar situation occurs at domain boundaries, where also $\tau_h \in O(1)$ as has been mentioned. However, in this case because it is difficult to determine a priori the required grid fineness at the boundary, local refinement is allowed and the number of refinements is limited by setting a maximum number of grids to be constructed.

## 5. SIMPLE / MULTILEVEL PROCEDURE FOR LOCALLY REFINED GRIDS

In this section the solution of the discrete system (3.13) on a fixed composite grid, denoted by the capital letter $H$, is considered. Lowercase letters such as $h$ will now refer to a particular *level* of the composite grid $H$. Also, since in this section the interest is not in the exact differential solution $u$, $v$, $p$ of (3.1), let now $\hat{u}_H, \hat{v}_H, \hat{p}_H$ be used instead of $u_h^*$, $v_h^*$, $p_h^*$ to denote the exact solution of the algebraic system (3.13), and let $u_H$, $v_H$, $p_H$ denote an estimate of this solution in the iterative solution procedure. With this notation the algebraic system (3.13) to be solved is written as:

$$N_H^x(\hat{u}_H, \hat{v}_H, \hat{p}_H) = 0 \quad , \quad N_H^y(\hat{u}_H, \hat{v}_H, \hat{p}_H) = 0 \quad , \quad N_H^c(\hat{u}_H, \hat{v}_H, \hat{p}_H) = 0 \qquad (5.1)$$

Suppose for simplicity that the composite grid $H$ consists of only 3 levels with global parts, $h$, $2h$ and $4h$, as depicted in Figure 5. During the multilevel cycle, the iterative procedure visits one by one all the levels of the grid and it solves on both the global *and* local part of each level. Therefore a grid function $\phi_h$ defined on a level $h$ will be written as $\phi_h = (\phi_h^l, \phi_h^g)$ where $\phi_h^l$, $\phi_h^g$ contain the values of $\phi_h$ at the local and global parts of level $h$ respectively. The exact solution may be written as $\hat{u}_H = (\hat{u}_h^g, \hat{u}_{2h}^g, \hat{u}_{4h}^g)$ etc. (of course it is defined only at global CVs). Now, suppose that at some point of the iterative procedure when the current estimate of the solution is $u_H^* = (u_h^{*g}, u_{2h}^{*g}, u_{4h}^{*g})$ etc. iterations pass from level $h$ to level $2h$. The current estimate $u_H^*$, $v_H^*$, $p_H^*$ satisfies (5.1) up to a residual. In particular, on level $h$ we would have that:

$$N_h^x\left(u_h^*,v_h^*,p_h^*\right)=-r_h^x \quad , \quad N_h^y\left(u_h^*,v_h^*,p_h^*\right)=-r_h^y \quad , \quad N_h^c\left(u_h^*,v_h^*,p_h^*\right)=-r_h^c \qquad (5.2)$$

(actually the left hand sides of (5.2) also involve variables of some of the CVs of level $2h$). If there were a still finer level then level $h$ would also have a local part, and apart from the subset of (5.1) which correspond to level $h$, (5.2) would also include equations for the local CVs, which would also be satisfied up to a residual. The form of these local equations will be described shortly.

Now, the iterative procedure moves to level $2h$. First the current solution is restricted from level $h$ to level $2h$ (including the current solution at the local part of level $h$) and stored as $\tilde{u}_{2h}^*=I_h^{2h}u_h^*$, $\tilde{v}_{2h}^*=I_h^{2h}v_h^*$, $\tilde{p}_{2h}^*=I_h^{2h}p_h^*$ (defined only on the local part of $2h$), where $I_h^{2h}$ is a restriction operator which is different from that of Section 4. The experience of the authors has shown that operator (5.3) is suitable also for high Reynolds number flows:

$$\left(I_h^{2h}\phi_h\right)_P = \frac{\sum_{C\in C_P}\phi_{h,C}\cdot\Delta\Omega_C}{\sum_{C\in C_P}\Delta\Omega_C} \qquad (5.3)$$

where $C_P$ is the set of 4 children (of level $h$) of CV $P$ (of level $2h$).

On level $2h$ the equations solved are such that their solution, say $u_{2h}, v_{2h}, p_{2h}$, is such that: On the global part of $2h$, the new solution $u_{2h}^g, v_{2h}^g, p_{2h}^g$ is a better estimate to $\hat{u}_{2h}^g, \hat{v}_{2h}^g, \hat{p}_{2h}^g$ than the previous estimate $u_{2h}^{*g}, v_{2h}^{*g}, p_{2h}^{*g}$. And on the local part of $2h$, the solution $u_{2h}^l, v_{2h}^l, p_{2h}^l$ produces corrections $u_{2h}^l-\tilde{u}_{2h}^*, v_{2h}^l-\tilde{v}_{2h}^*, p_{2h}^l-\tilde{p}_{2h}^*$ which when prolonged back to level $h$ give an estimate which satisfies the equations of level $h$ more closely than the previous estimate $u_h^*, v_h^*, p_h^*$. The equations of level $2h$ are the following:

*Global CVs with only global neighbours* (CVs indicated as ● and ○ in Figure 5): At these CVs the equations solved are precisely those of the composite grid, i.e. the corresponding subset of (5.1). For CVs ○, the variables at their coarse neighbours + are not treated as unknowns but as fixed Dirichlet boundary values ($u_{4h}^{*g}, v_{4h}^{*g}, p_{4h}^{*g}$). The discretisation stencils of CVs ○ also involve the gradients at CVs +, which in turn involve the variables at CVs ○; therefore in our procedure we have chosen to update the values of the gradients at CVs + after each iteration on level $2h$. Concerning the gradients, it should also be mentioned that because the equations of some of these global CVs also make use of the gradients at global CVs of type □, the gradients at CVs □ must be calculated grid-wise (like in Figure 3) and not level-wise, i.e. assuming that the neighbours of the CVs □ are the fine CVs × and not the local CVs ■. For the calculation of these gradients in the present work the values at CVs × were assumed to be fixed at $u_h^{*g}, v_h^{*g}, p_h^{*g}$ rather than updating them after each iteration according to the corrections produced at CVs ■.

*Local CVs* (CVs indicated as ■ in Figure 5): First we consider the momentum equations. Suppose a local CV $P$ of level $2h$, and $C_P$ the set of its children for which equations of the type (5.2) hold. According to the multigrid methodology (see e.g. [5] or [27]) if the iteration errors and the residuals per unit volume which correspond to the estimate $u_h^*, v_h^*, p_h^*$ are smooth enough compared to the spacing of level $h$, then the subset of 4 $x$-momentum equations of (5.2) which correspond to the children of $P$ may be approximated by a single $x$-momentum equation of CV $P$ as:

$$N_{2h,P}^x\left(u_{2h},v_{2h},p_{2h}\right) = N_{2h,P}^x\left(\tilde{u}_{2h}^*,\tilde{v}_{2h}^*,\tilde{p}_{2h}^*\right) + \sum_{C\in C_P}r_{h,C}^x \qquad (5.4)$$

Equations (5.4) are defined only at the local part of 2*h*. The operator $N_{2h}$ need not be the same as $\mathcal{N}_{2h}$, but it must approximate the same exact integral-differential operator $N$ (3.1). If $P$ has a global neighbour then the left hand side of (5.4) depends also on $u_{2h}^g, v_{2h}^g, p_{2h}^g$, and the first term of the right hand side depends also on $u_{2h}^{*g}, v_{2h}^{*g}, p_{2h}^{*g}$. Therefore if the equations of the global CVs are not satisfied then the solution $u_{2h}$, $v_{2h}$, $p_{2h}$ of (5.4) will be different from $\tilde{u}_{2h}^*, \tilde{v}_{2h}^*, \tilde{p}_{2h}^*$ even if $r_h = 0$. The *y*-momentum equations are treated in the same way.

Now we turn to the continuity equation. Normally the continuity equation of local CVs should be defined as (5.4) with superscript *x* replaced by *c*. However, following the usual SIMPLE/multigrid algorithm, we depart slightly from this rule and define the local continuity equation similarly to the global continuity equation, as:

$$\sum_{f \in f_P} F_{2h,f} = \sum_{f \in f_P} \left\{ \tilde{F}_{2h,f} + \left[ F_{2h,f}\left(u_{2h}, v_{2h}, p_{2h}\right) - F_{2h,f}\left(\tilde{u}_{2h}^*, \tilde{v}_{2h}^*, \tilde{p}_{2h}^*\right) \right] \right\} = 0 \quad (5.5)$$

where:

$$\tilde{F}_{2h,f} = \sum_{c \in c_f} F_{h,c}\left(u_h^*, v_h^*, p_h^*\right) \quad (5.6)$$

In the above, $f_P$ is the set of 4 local faces of the local CV *P*, and $c_f$ is the set of 2 child faces of the local face *f*. In (5.5), $F_{2h,f}$ are defined by (3.8). If *f* separates *P* from a global CV then for the calculation of $F_{2h,f}(\tilde{u}_{2h}^*, \tilde{v}_{2h}^*, \tilde{p}_{2h}^*)$ in (5.5) and $F_{h,c}$ in (5.6) values from $u_{2h}^{*g}, v_{2h}^{*g}, p_{2h}^{*g}$ will also be used. In (5.6), $F_{h,c}$ is to be replaced by $\mathcal{F}_{h,c}$ if the child *c* is itself local. In other words, the mass flux $\tilde{F}_{2h,f}$ is defined as the sum of the mass fluxes through the children of *f* at the time of the restriction from level *h* to level 2*h*; and the mass flux $F_{2h,f}$ is defined as equal to the restricted mass flux $\tilde{F}_{2h,f}$ plus a correction (the term in square brackets in (5.5)) which is due to the improvement of the flow field estimate. A slight improvement of this definition of the local mass fluxes will be proposed in Section 6. Summarising, the local continuity equation is defined as $\sum F_{2h,f} = 0$, like its global counterpart, but the mass fluxes though local faces, $\mathcal{F}_{2h}$, are defined differently than $F_{2h}$, using (5.5) instead of (3.8). This is necessary so that the local continuity equations will produce corrections which are driven by the residuals of the finer level. In fact, by summing the continuity equations of the children of *P*, it is not hard to see from definition (5.6) that:

$$\sum_{f \in f_P} \tilde{F}_{2h,f} = -\sum_{C \in C_P} r_{h,C}^c \quad (5.7)$$

Using (5.7), equation (5.5) becomes:

$$\sum_{f \in f_P} F_{2h,f}\left(u_{2h}, v_{2h}, p_{2h}\right) = \sum_{f \in f_P} F_{2h,f}\left(\tilde{u}_{2h}^*, \tilde{v}_{2h}^*, \tilde{p}_{2h}^*\right) + \sum_{C \in C_P} r_{h,C}^c \quad (5.8)$$

which is precisely (5.4) with superscript *x* replaced by *c*. So in fact it is not the continuity operator which is defined differently on local and global CVs, but the momentum operator because $N_{2h}^x$ uses the mass fluxes $F_{2h}$ while $\mathcal{N}_{2h}^x$ uses the mass fluxes $\mathcal{F}_{2h}$.

*Global CVs with local neighbours* (CVs indicated as □ in Figure 5): This is the most difficult case. The equations must involve unknowns of the same level, and therefore the unknowns at CVs ■. However, at the overall convergence of the multigrid procedure the composite grid equations must be satisfied, which involve the global unknowns at CVs × of level *h* and not the local unknowns at CVs ■ of level 2*h*. This is easy to achieve as far as the continuity equation is concerned. Indeed, if *P* is a CV of type □ and $g_P$, $s_P$ are the sets of global and local (sub-exterior) faces of *P* (regarding *P* in a level-wise manner, bounded by 4 faces) then its continuity equation has the natural form:

$$\sum_{f \in g_P} F_{2h,f}\left(u_{2h}, v_{2h}, p_{2h}\right) + \sum_{s \in s_P} F_{2h,s}\left(u_{2h}, v_{2h}, p_{2h}\right) = 0 \tag{5.9}$$

At overall convergence of the multigrid cycles the "correction" term in square brackets in (5.5) of the local mass fluxes becomes zero and $F_{2h} = \tilde{F}_{2h}$. Substituting this into (5.9) and using (5.6), (5.9) becomes:

$$\sum_{f \in g_P} F_{2h,f}\left(u_{2h}, v_{2h}, p_{2h}\right) + \sum_{s \in s_P} \sum_{c \in c_s} F_{h,c}\left(u_h, v_h, p_h\right) = 0 \tag{5.10}$$

($F_{h,c}$ also involve values from $u_{2h}$, $v_{2h}$, $p_{2h}$) which is precisely the composite grid equation, i.e. the $P$-th continuity equation of the system (5.1), as is required.

A similar reasoning lies behind the construction of the momentum equations. Let the net $x$-momentum fluxes (3.7) be written as $M^x_{h,f}[(u^1_*, u^2_*),(v^1_*, v^2_*),(p^1_*, p^2_*)]$ to indicate that face $f$ separates two CVs and the variables of the first of these CVs belong to the arrays $u^1_*, v^1_*, p^1_*$ while the variables of the second CV belong to $u^2_*, v^2_*, p^2_*$. Then, to simplify the descriptions which follow, define the following abbreviations, where $e, f, s$ denote respectively an exterior face of level $h$, a global face of level $2h$, and a local sub-exterior face of level $2h$:

$$\begin{aligned}
M^{x,*}_{h,e} &\equiv M^x_{h,e}\left[\left(u^{*g}_{2h}, u^{*g}_h\right),\left(v^{*g}_{2h}, v^{*g}_h\right),\left(p^{*g}_{2h}, p^{*g}_h\right)\right] \\
M^x_{2h,f} &\equiv M^x_{2h,f}\left[\left(u^g_{2h}, u^g_{2h}\right),\left(v^g_{2h}, v^g_{2h}\right),\left(p^g_{2h}, p^g_{2h}\right)\right] \\
M^x_{2h,s} &\equiv M^x_{2h,s}\left[\left(u^g_{2h}, u^l_{2h}\right),\left(v^g_{2h}, v^l_{2h}\right),\left(p^g_{2h}, p^l_{2h}\right)\right] \\
M^{x,*}_{2h,s} &\equiv M^x_{2h,s}\left[\left(u^{*g}_{2h}, \tilde{u}^*_{2h}\right),\left(v^{*g}_{2h}, \tilde{v}^*_{2h}\right),\left(p^{*g}_{2h}, \tilde{p}^*_{2h}\right)\right]
\end{aligned} \tag{5.11}$$

Thus $M^{x,*}_{h,e}$ is the estimate of the net $x$-momentum flux through face $e$ at the time of restriction from level $h$ to level $2h$; $M^x_{2h,f}$ and $M^x_{2h,s}$ are the fluxes through $f$ and $s$ which result from the solution of the system of (both local and global) equations of level $2h$; and $M^{x,*}_{2h,s}$ is the flux through $s$ which is calculated immediately after restriction from level $h$ to level $2h$.

Then, with reasoning similar to that for the continuity equation, the $x$-momentum equation for a CV $P$ of type □ is defined as:

$$\sum_{f \in g_P} M^x_{2h,f} + \sum_{s \in s_P}\left[\sum_{e \in c_s} M^{x,*}_{h,e} + \left(M^x_{2h,s} - M^{x,*}_{2h,s}\right)\right] = 0 \tag{5.12}$$

That is, the momentum flux through each local face $s$ of $P$ is defined as equal to the sum of $M^{x,*}_{h,e}$, the estimates of the fluxes through the exterior children of $s$ at the time of restriction, plus a correction term (in parentheses in (5.12)) which is due to the improvement of the flow field estimate. At convergence this correction term is zero, and what remains is precisely the composite grid equation, the $P$-th $x$-momentum equation of the system (5.1).

This completes the definition of the equations at all CVs of level $2h$. After the required number of iterations have been performed at level $2h$, iteration may pass to a still coarser level $4h$ etc. At some point of the multilevel cycle, iteration will return to level $2h$ and then it will move up to level $h$. At this point prolongation of the corrections produced at the local part of $2h$ will occur, to give an improved estimate of the solution of the equations of level $h$:

$$u_h \leftarrow u_h^* + a_{mg} I_{2h}^h \left( u_{2h}^l - \tilde{u}_{2h}^* \right) \tag{5.13}$$

and similarly for $v$, $p$. In (5.13) $a_{mg}$ is a real number, usually $a_{mg}=1$ but a smaller number may be used if convergence problems occur, and $I_{2h}^h$ is a prolongation operator, which in the present work is defined as:

$$\left( I_{2h}^h \phi_{2h} \right)_C = \phi_{2h,P} + \left( \nabla_{2h} \phi_{2h} \right)_P \cdot (C - P) \tag{5.14}$$

where $P$ is the parent of $C$, and $P$, $C$ are their centres.

The initial guess for the solution of the equations of a particular level, say $2h$, is $\tilde{u}_{2h}^*, \tilde{v}_{2h}^*, \tilde{p}_{2h}^*$ at the local part and $u_{2h}^*, v_{2h}^*, p_{2h}^*$ at the global part. Then it is not difficult to see that if the present estimate $u_H$, $v_H$, $p_H$ equals the exact solution $\hat{u}_H, \hat{v}_H, \hat{p}_H$ of (5.1), then the initial guess at any level satisfies the equations of that particular level (both local and global) and zero corrections are produced at the local parts. In other words, the multilevel cycle does not alter the exact solution.

Since the adopted discretisation uses the central difference scheme (3.5) (CDS), the solution of the system of equations of any particular level may exhibit oscillations, as is discussed in [25]. The problem of pressure oscillations is taken care of by the use of momentum interpolation for the mass fluxes in both local and global equations of each level – see [25] (however, see Section 6 for further discussion). Some care is needed to avoid velocity oscillations which may appear especially at coarse levels and high Reynolds numbers. For the multilevel procedure, the problem is twofold: First, during the restriction phase, say from level $h$ to level $2h$, any velocity oscillations in the field $u_h^*, v_h^*, p_h^*$ may reflect in a restricted field $\tilde{u}_{2h}^*, \tilde{v}_{2h}^*, \tilde{p}_{2h}^*$ which causes the multilevel procedure to fail to converge. This has indeed been observed in practice at high Reynolds numbers when other restriction operators are used, but it seems that the operator (5.3) overcomes this problem. Conversely, during the prolongation of the corrections from level $2h$ to level $h$, if the solution $u_{2h}$, $v_{2h}$ contains oscillations then the corrections will also be oscillatory. These oscillations will appear to have a greater wavelength compared to the mesh spacing on level $h$ than on level $2h$. In fact, since at any level the equations are not solved exactly but only a few iterations are performed, these oscillations will survive until prolongation to the finest levels. On these finest levels the wavelength of the oscillations will appear to be very large compared to the mesh spacing, and therefore the smoother will be unable to reduce them. The impact on the overall efficiency of the multilevel procedure will be detrimental. One known solution to this problem (see [2]) is to define the local momentum fluxes using a blend of the CDS with the 1st order upwind scheme (UDS) for the convection terms. That is, replace $F_{h,f} \cdot u_{h,c}$ by $F_{h,f} \cdot [a_c \cdot u_{h,c} + (1-a_c) \cdot u_{h,c}^{UDS}]$ in (3.7), where $0 \leq a_c < 1$ and $u_{h,c}^{UDS}$ is the value of $u$ at the centre of the face as given by the 1st-order UDS scheme (it is equal to the value at the centre of the adjacent CV from which fluid flows towards the face). The global fluxes remain unaltered so that the composite grid equations (5.1) still use pure CDS. This change results in a smoother solution within the local part of each level, but because the local operator $N_{2h}$ departs from $N_h$, the corrections produced are less effective and the convergence rate of the multilevel procedure drops. This will be shown in Section 6. Alternatively, it has been found that it is more efficient to leave the discretisation of the local momentum fluxes as it is and to smooth the corrections prior to prolongation using a smoothing operator $S_{2h}$, so instead of (5.13) prolongation takes place as follows:

$$u_h \leftarrow u_h^* + a_{mg} I_{2h}^h S_{2h} \left( u_{2h}^l - \tilde{u}_{2h}^* \right) \tag{5.15}$$

The corrections of $v$ are similarly smoothed, while the corrections of $p$ need not be smoothed. The smoothing operator was inspired by the fact that, as discussed in [25], in the absence of momentum interpolation an oscillatory pressure field results, but which results in near-correct values of pressure at face centres. Therefore, $S_h$ is defined as:

$$\left(S_h \phi_h\right)_P = \frac{1}{4} \sum_{f \in f_P} \overline{\phi}_{h,c'} \tag{5.16}$$

where $f_P$ is the set of 4 faces of CV $P$ (in level-wise treatment every CV has exactly 4 faces) and the overbar denotes linear interpolation at point $c'$ of face $f$ from the values at adjacent CV centres. The efficiency of this technique will be demonstrated in Section 6.

Any smoother can be used in the context of the multilevel procedure as described so far. Now some particular issues which concern the use of SIMPLE will be addressed. First of all, a problem arises in the presence of outlet boundaries, when the outlet boundary condition is implemented in the usual way (e.g. as in the CAFFA code provided with [2] – see also [27]). In this implementation, after solution of the velocity linear systems within each SIMPLE outer iteration the outlet mass fluxes are scaled to satisfy overall mass conservation through the boundaries of the domain. This actually interferes with the discretisation of the equations and produces a final solution which does not exactly fulfil the zero-gradient condition at the outlet – see [27] for more details. The problem with the multilevel procedure is that since a particular level may not include all global outlet faces, it is difficult to find a suitable way to update the outlet mass fluxes of the particular level while simultaneously ensuring global mass conservation. To overcome this problem, it was chosen in this work not to alter the outlet mass fluxes during SIMPLE smoothing sweeps on any particular level, but to perform global SIMPLE sweeps on the whole composite grid between multilevel cycles. These will be termed *composite-grid smoothing sweeps* in contrast to the *level smoothing sweeps* which occur within each multilevel cycle.

Another complication arises at the exterior CVs of each level. Because the coarse CVs + (Figure 5) do not contribute unknowns to the system of equations of level $2h$, it so happens (see [27]) that the exterior faces of $2h$ may have negative contributions to the main diagonal of the matrix of the velocity linear systems of each SIMPLE sweep on level $2h$. Thus for the solution of these systems we used unpreconditioned GMRES which does not require strict diagonal dominance. It was observed that usually 4-5 inner iterations suffice to achieve the full rate of convergence of outer iterations. Also, for the construction of the pressure correction system of SIMPLE, the mass flux correction through a face assumes a form which is a function of the main diagonal coefficients of the velocity linear systems of the CVs which lie on either side of the face. For exterior faces, during level sweeps, the coefficient of the fine CV is unsuitable as mentioned above, and the coefficient of the coarse CV does not exist. However, since composite-grid smoothing sweeps occur between multilevel cycles, we store the contributions of exterior faces to the matrix of coefficients of the pressure correction system during each composite-grid sweep and use them within the immediately following cycle. Since at each level sweep the exterior face mass fluxes are also corrected, the pressure correction system becomes diagonally dominant and always has a single solution (except if the particular level does not have exterior faces). We use ILU(0)-preconditioned conjugate gradients for the pressure correction system.

Finally, the matter of the type of multilevel cycle must be briefly addressed. For structured grids it has been found that for medium and high Reynolds numbers W cycles are usually more efficient than V cycles. However, for composite grids this may not be the case if the finest levels have very few CVs. In such cases it is often more efficient to use a cycle which resembles a W cycle at the lower levels and a V cycle at the higher levels, like the one shown in Figure 6. In the following, the notation $W^k/V(v_1,v_2)$–$s$ means a multilevel procedure where on levels up to the $k$-th 2 cycles are used to solve the problem of the immediately finer level, while 1 cycle is used on levels $> k$. Also $v_1$, $v_2$ are the number of pre-smoothing and post-smoothing iterations, and $s$ is the number of composite-grid sweeps between multilevel cycles.

## 6. NUMERICAL EXPERIMENTS

### *6.1 Square Lid-Driven Cavity*

The square lid-driven cavity problem is probably the most often used case to test new methods for incompressible flows. It is used in many of the papers mentioned in Section 1 ([6], [7], [12], [13], [15], [16], [18], [19]). Usually methods are validated by comparing their results against those of [30], but these results are not very accurate. For the particular case of Re=1000 the very accurate results of [31] are available. In the following, the cavity has side L = 1 m while the lid velocity and fluid density are fixed at V = 1 m/s (in the positive *x*-direction) and $\rho$ = 1 kg/m$^3$. The viscosity $\mu$ is varied according to the required Reynolds number Re = $\rho$VL/$\mu$. The quantities $Q_{mom}$ and $Q_{mas}$ are defined as $Q_{mom}$ = $\rho$LV$^2$ = 1 kg·m/s$^2$ and $Q_{mas}$ = $\rho$LV = 1 kg/s.

For comparison, Table 1 contains descriptions of the convergence of the classic SIMPLE/multigrid method ([8]) for various cases. Uniform grids with square CVs are used. For each case the table displays two numbers: The number of cycles required to reduce the maximum residuals per unit volume of all equations below 10$^{-8}$ is displayed on the left. On the right (in *italic*) the corresponding residual reduction factor defined as:

$$q = \sqrt[m]{\frac{\left\|R_h^{m_0+m}\right\|_\infty}{\left\|R_h^{m_0}\right\|_\infty}} \qquad (6.1)$$

is given, where $R_{h,P}^k = r_{h,P}^k/\Delta\Omega_P$ is the residual per unit volume after cycle *k*, and *m*, $m_0$ are large enough so that the rate of convergence has stabilised but *q* is not influenced by the first $m_0$ cycles where the rate of convergence may be irregular. The maximum value over the momentum and continuity equations is displayed (usually the reduction factors are nearly equal for the three sets of equations). Unless otherwise indicated Full Multigrid (FMG) is used with parameters $a_u$=0.8, $a_p$=0.2 (SIMPLE underrelaxation factors), $a_{mg}$=1, $a_{mi}$=1, and the coarsest level (CL) is level 1 (8×8 CVs). For low Re numbers V cycles are more efficient, while W cycles are more efficient for medium and high Re numbers. For Re $\geq$ 5000 velocity oscillations make it impossible to attain convergence unless either the velocity corrections are smoothed according to (5.15) or a blend of CDS/UDS is used for the local convective momentum fluxes. We have found that it is always more efficient to smooth the corrections, as demonstrated in Figure 7, and it is this technique which was used to obtain the results for Re=5000, 10000 in the Tables.

Table 2 shows the same data for the new SIMPLE/multilevel method described in Section 5. For low and medium Re the results are similar, but the method becomes very inefficient at high Re numbers. In fact, it is not possible to attain convergence unless $a_{mi}$ is dropped to 0.5 for Re=2000 (not shown) and 0.1 for Re=5000, resulting in great inefficiency. The larger the distance between the finest and the coarsest level, the smaller $a_{mi}$ must be. For example, Table 2 shows that by letting the coarsest level CL=2 (16×16 CVs) it is possible to increase $a_{mi}$ to 0.2, increasing the efficiency. The problem seems associated with momentum interpolation, and an outline of a possible explanation is that it is caused by the fact that local mass fluxes (5.5) contain not one but two separate pressure terms. The fixed pressure term of $F_{2h,f}(\tilde{u}_{2h}^*, \tilde{v}_{2h}^*, \tilde{p}_{2h}^*)$ may be relatively large, especially at coarse levels because it is calculated from the restricted pressure field, and it must be balanced by the "dynamic" pressure term of $F_{2h,f}(u_{2h},v_{2h},p_{2h})$ and by the mass flux change which is due to the velocity correction. This spoils the corrections to be prolonged back to the finer levels. On the other hand, in the classic SIMPLE/multigrid method the local mass fluxes contain a single pressure term like global mass fluxes, and its magnitude should in general be negligible compared to the total flux (see [25]). Now, if $F_{2h,f}(\tilde{u}_{2h}^*, \tilde{v}_{2h}^*, \tilde{p}_{2h}^*)$ in (5.5) was defined with $A_f(u_{2h},v_{2h})$ in (3.8) instead of $A_f(\tilde{u}_{2h}^*, \tilde{v}_{2h}^*)$ then the sum of the two pressure terms would equal the single pressure term of the classical SIMPLE/multigrid method involving the pressure correction, and the two methods would become equivalent. However this would require that $F_{2h,f}(\tilde{u}_{2h}^*, \tilde{v}_{2h}^*, \tilde{p}_{2h}^*)$ be re-calculated at every smoothing sweep (because the value of $A_f(u_{2h}, v_{2h})$ would change) and to avoid this it was chosen instead to replace $A_f(u_{2h}, v_{2h})$ by $A_f(\tilde{u}_{2h}^*, \tilde{v}_{2h}^*)$ in the definition (3.8) of $F_{2h,f}(u_{2h}, v_{2h}, p_{2h})$ of (5.5). This again makes the sum of the two pressure terms equivalent to a single pressure term involving the pressure correction, but it is not equivalent to the pressure term of the classic method. The results shown in Table 3 were obtained with this method and it is clear that the situation has improved, with $a_{mi}=1$ used throughout. In the case Re=10000 $a_{mg}=0.8$ had to be used for the velocity, and efficiency is somewhat lower than that of the classic method, but still it is considered adequate. This modification will be used in all subsequent experiments. It should be noted here that the difference between $A_f(u_{2h}, v_{2h})$ and $A_f(\tilde{u}_{2h}^*, \tilde{v}_{2h}^*)$ becomes greater as Re increases as seen from (3.10) (because their convective parts are different but their viscous parts are the same). This may explain why the problem occurred at high Re. The table also shows the CPU time required by the FMG procedure on a 1.4 GHz Pentium 4 processor. The time required per cycle for the results of tables 1 and 2 is nearly the same as for table 3.

The next step is to allow local refinement. Starting from level 4 (64×64) local refinement levels are added according to the criterion (4.2). At the upper corners of the cavity the flow field is discontinuous and this causes the truncation error to actually increase near the corners as the grid is refined. Therefore, the refinement criterion cannot be satisfied and the process of refinement must be ended by setting a maximum allowable number of grids. For Re = 100 and 5000, $r_\tau = 0.01$ was used and two additional grids were allowed (with level 6 being the finest). The underlying grids of the corresponding final grids are shown in Figure 8. Local refinement occurs at the regions where the estimate (4.1) predicts a high truncation error, that is near the top lid and corners for Re = 100 and at the circumference of the main vortex for Re = 5000 (see [27]). The number of cycles required to reduce the maximum residuals per unit volume below $10^{-8}$ and the corresponding reduction factors are shown in Table 4, where grid 1 comes from grid 64×64 with local refinement, and grid 2 (whose underlying grid is shown in Figure 8) comes from grid 1. For each grid the finest level is written in parentheses. The FMG procedure is used in each case starting from grid 8×8. For Re = 100 V(2,2)-1 cycles were used and comparison with Table 3 shows that the number of cycles and the reduction factors are the same as for the non-composite grids. For Re = 5000 $W^k$/V(2,2)-1 cycles were used with different values of $k$ as shown in the Table. In this case the highest possible value $k=5$ proves to be the most efficient, also in terms of CPU time. A slight degradation of the convergence rate is observed for grid 2 compared to grid 256×256.

For Re = 1000 $r_\tau = 0.001$ was used and four additional grids were allowed (level 8 being the finest) to study the benefits of local refinement comparing with the accurate results of [31]. The underlying grids of the resulting series of 4 grids are shown in Figure 9. As expected, because the refinement criterion is dynamic, not only is a new level added on each subsequent grid but also existing levels are extended. The truncation error is highest at the lid and at the circumference of the main vortex, especially at its right and lower part. Figure 10 shows the $u$-velocity discretisation error $|\varepsilon_h^u|$ ($h$ again denotes the whole grid, not a particular level) along the vertical centreline on various composite and non-composite grids, calculated at the points and from the $u$- values given in [31], which are regarded as "exact". The distance between the distributions of the non-composite grids is in accordance with the 2$^{nd}$ order accuracy of the method. Composite grid 1 and grid 128×128 have about the same number of CVs and offer nearly identical accuracy. The accuracy offered by composite grid 2 is very close to that of grid 256×256 although it has half as many CVs. Composite grid 3 has about the same number of CVs as grid 256×256 but offers a clearly more accurate solution. The accuracy of composite grid 4 is comparable to that of the much larger (in terms of number of CVs) grid 512×512, except near the centre of the cavity where however the error of the 512×512 solution is already very small and so this difference is not so significant. The addition of a relatively large number of CVs to get grid 4 from grid 3 does not seem to reflect in an equivalent reduction in error. To interpret this one must keep in mind that the new CVs are mostly located near the lid corners (see Figure 9), where the increase of the grid density is not so efficient in improving the accuracy due to the singularities. Also, the introduction of many refinement levels increases the area of the domain which is near level interfaces, where there are large $O(1)$ truncation errors as mentioned in Section 3. Already this area is comparable to the area occupied by the whole of the finest level of grid 4. Finally, according to Tables 3 and 4 the convergence rates of SIMPLE/multilevel are similar to those on the non-composite grids but convergence is achieved in less cycles because successive composite grids differ less than successive non-composite grids (since the finest levels cover only a small proportion of the domain) and so the initial guess due to FMG has a smaller error on the composite grids. This time using cycles $W^k/V$ with an increased number $k$ does not seem to pay off, since the finest levels cover less space, and it is more efficient to use $k = 3$ or 4.

Next the procedure was tested on non-uniform grids. Figure 11 (left) shows a non-uniform 32×32 grid, where the ratio of the heights or the widths of consecutive CVs is $r_{32} = 1.1561$. Again for Re = 1000, using $r_\tau = 0.05$ and allowing 3 additional grids produces Grid 1 shown in Figure 11 (right). Then Grid 2 is produced by refining every CV of Grid 1, and Grid 3 is produced by refining every CV of Grid 2. The number of global CVs of each level of Grid 1 and the percentage of the domain that they occupy are displayed in Table 5. Also, for comparison a series of non-composite grids are constructed, up to 256×256, which are such that $r_{64} = \sqrt{r_{32}}$, $r_{128} = \sqrt{r_{64}}$ etc. The grid lines of any given grid are also grid lines of the immediately finer grid. Figure 12 again displays $|\varepsilon_h^u|$ along the vertical centreline on various grids. Again the distance between the distributions of successive non-composite grids is in accordance with the 2$^{nd}$-order accuracy of the method. However, the distance between the distributions of the composite grids also suggests 2$^{nd}$-order accuracy. This is a very interesting observation in view of the fact that the truncation error has magnitude of $O(1)$ near level interfaces, as noted in Section 3. Unfortunately in [31] only results along the centrelines are given, so it can not be strictly verified that convergence is 2$^{nd}$-order throughout the domain. Of course, 2$^{nd}$ order accuracy with respect to refinement of the whole grid does not mean that the addition of any single level may not even cause an increase of the discretisation error due to the truncation error increase at the new level interface. Comparing the error distributions of Figure 12, there does not appear to be any clear benefit in using the composite rather than the non-composite grids. A possible explanation is that the non-composite grids are already nearly optimal. This can be seen also from the fact that the two finest levels cover very small percentages of the domain. These finest levels are located near the top corners where the flow field is singular which limits their contribution towards the increase of accuracy. Also, despite the fact that they cover only a very small percentage of the domain, they consist of a relatively large number of CVs due to the non-uniformity of the grid. Also, in consecutive non-composite grids the maximum width to height ratio of the CVs increases (from 8.8 in 32×32 to 10 in 256×256) while this ratio remains constant at 8.8 in the composite grids because the child CVs inherit it from their parent during refinement. This means that the spacing of the non-composite grids near the walls is smaller than that of the composite grids.

Figure 13 shows the estimate $\tau_h^{x*}$ (4.1) near the top right corner on grids 2 and 3, which is seen to predict that grid refinement causes the area of high $\tau_h^x$ to reduce in size, and the maximum $\tau_h^x$ to increase. The estimate is also seen to predict high $\tau_h^x$ at the region covered by the CVs of $V_{2h}^e \cup V_{2h}^{en}$, which is due to the fact that the estimate (4.1) uses the underlying grid $2h$ to calculate the truncation error, while actually $\tau_h^x \in O(1)$ at the region covered by the CVs of $V_h^e \cup V_h^{en}$. Similarly, the estimate shows high truncation error in a region near the boundaries which is twice as wide as the actual one.

Table 6 displays the number of cycles required to drop the maximum residual per unit volume below $10^{-8}$ (FMG is used) and the residual reduction factors for various cases. The reduction factors are somewhat worse than for the uniform grids, and this is due to the aspect ratio of the CVs (it is a well known problem that multigrid smoothers loose their efficiency when the grid spacing in one direction is much smaller than in the other – see [5]). For the composite grids only the results for the most efficient type of cycle (in terms of $k$) are displayed, in terms of CPU time (level 32×32 is level 3). It was observed that in general increasing the number of $k$ causes the number of cycles and the reduction factor to drop, but since it also causes the cycle to become more expensive it does not pay off to increase $k$ as far as possible.

*6.2 Skew lid-driven cavities*

Next the procedure is tested on a problem which is often used for testing on non-orthogonal grids, the flow in a cavity whose side walls are inclined at an angle of β=45° or β=30° to the horizontal level. This problem is proposed in [32]. All sides of the cavity have a length of L = 1 m and the flow and solution parameters are the same as in Section 6.1 unless otherwise stated. Starting from a uniform 64×64 grid (level 4) two additional grids are allowed according to a criterion $r_\tau$ = 0.001. Simulations are performed for Re = 100 and Re = 1000. The underlying grids of the resulting final grids are shown in Figure 14. Again for Re = 100 the truncation error is higher near the lid and top corners, while for Re = 1000 this also occurs at the circumference of the main vortex which is smaller in size than that of the square cavity and is located near the top right corner. Figure 15 shows the $u$ discretisation error distributions along the centrelines of the cavities which are parallel to the side walls. To calculate the discretisation error the exact $u$ values were estimated with Richardson extrapolation (see [2] or [22]), using the solutions of the 128×128 and 256×256 grids and assuming 2$^{nd}$ order accuracy. This was done because it was observed that the results presented in [32] are not significantly more accurate than the 256×256 solution. For Re = 100, the benefits of using local refinement are not evident, as the solution on the final composite grid is of comparable accuracy as on grid 128×128 and the two grids also have a comparable number of CVs. On the other hand it seems advantageous to use composite grids for Re = 1000: in regions of high discretisation error the solutions on composite grids 1 and 2 have nearly the same accuracy as the solutions on grids 128×128 and 256×256 respectively, although they have significantly less CVs. It is not surprising that local refinement works better with higher Reynolds numbers: Convection tends to transport discretisation errors generated at regions of high truncation error to distant locations without alteration, while diffusion (viscous forces) tends to transport discretisation errors with an ever decreasing magnitude as the distance from the source increases. Therefore, it pays off more to use refinement to reduce locally high truncation errors in the presence of strong convection.

The convergence of the SIMPLE/multilevel procedure, again for a criterion of 10$^{-8}$, is shown in Tables 7 and 8, for the non-composite and composite grids respectively. Level 1 (8×8) is used as the coarsest level in the FMG solution procedure. Each outer SIMPLE iteration includes a 2$^{nd}$ pressure correction step to account for grid non-orthogonality as suggested in [2]. For β=45° convergence is similar as for the square cavity, but for β = 30° some additional difficulties arise. For Re = 100 V(2,2) cycles no longer converge and either W(2,2) or V(3,3) have to be used. Also on composite grids for β = 30° it pays to increase the number of composite-grid SIMPLE sweeps to 4. Except for the case β=30° Re=100, which does not exhibit typical multigrid convergence, in the other cases the convergence rates on the composite grids are about the same as on the non-composite grids. The gains of the local refinement procedure in terms of CPU time are not as high as in terms of number of CVs. This may be due to the compiler used. The difference is that our code allocates memory for the CVs of the non-composite grids all at once (in the form of arrays), while the CVs of local refinement levels are allocated one at a time to allow for unrefinement (a feature not used in the present work).

*6.3 Backward facing step*

Finally, a brief description of the solution of the backward facing step problem (see e.g. [22], [33], [34]) is provided to test the method on problems with outlet boundaries. The domain and level 1 are shown in Figure 16. Like in the aforementioned studies, the narrow channel before the step is not included as part of the computational domain, but a fully developed velocity profile is assumed at the height of the step ($x=0$). More specifically, the top and bottom boundaries are solid walls, the right boundary is the outlet, while the left boundary (step) is a solid wall from $y = -0.5$ to $y = 0$ and an inlet boundary from $y = 0$ to $y = 0.5$ with a parabolic velocity profile $u = 24y(0.5-y)$U, U = 1 m/s. The grid is uniform with square CVs up to the middle of the domain ($x = 15$) and afterwards it becomes stretched under a mild constant stretching factor. Simulations were performed for Reynolds numbers Re = $\rho$UH/$\mu$ of 133 (low), 400 (medium) and 800 (high) (transition starts at Re $\approx$ 1150). Again $\rho$ = 1 kg/m$^3$ and $\mu$ is varied according to the Reynolds number.

Tables 9 and 10 contain convergence data for non-composite and composite grids respectively. Again FMG is used and the algebraic convergence criterion is $10^{-8}$. Level 1 is the coarsest level of the multilevel procedure. The composite grids are constructed starting from the 440×24 grid (level 3) and using $Q_{mom}$ = 1 kg·m/s$^2$, $Q_{mas}$ = 1 kg/s, $r_\tau$ = 0.01, allowing two additional grids. This time it is useful to perform smoothing of corrections according to (5.15) for the whole range of Reynolds numbers, and the results of Tables 9 and 10 include such smoothing. It is observed that, in general, increasing the grid fineness causes the reduction factor to increase, contrary to the ideal multigrid properties. The situation improves significantly if the number of composite-grid smoothing sweeps is increased. This makes likely the following explanation: Since outlet mass fluxes and velocities are not updated during each multilevel cycle, this means that the composite grid sweeps are totally responsible for the reduction of the errors of these outlet mass fluxes and velocities. These errors also contain smooth components, which would normally be reduced on coarse levels in a multilevel procedure, but now they have to be reduced by the composite grid SIMPLE sweeps. Like most single-grid solvers, SIMPLE becomes less efficient as the grid density increases which reflects in increased overall reduction factors in Tables 9 and 10. Fortunately, it seems that a significant improvement of the convergence rate results with only a small increase in the number of composite-grid sweeps.

## 7. CONCLUSIONS

A multilevel algorithm has been proposed for locally refined grids, and tested using SIMPLE as the smoother. The tests show that the existence of local refinement levels does not adversely affect the convergence rates, which are similar to those of the classic SIMPLE/multigrid method. Contrary to the algorithm of [20], the present algorithm does not require the partitioning of the domain into blocks, and solution takes place on the whole composite grid simultaneously and not block-by-block. Therefore, one expects that the present algorithm is more efficient than that of [20]. However, the algorithm of [20] is more easily parallelizable. Of course, the two approaches can be combined by partitioning the domain into blocks each of which is itself a composite grid.

Although the main focus of this work has been on the multilevel solver, several issues related to the local refinement criterion itself have risen:

Local refinement was driven by a criterion which tries to minimise the integral of the truncation error by refining the CVs which contribute the greatest to this integral. The tests have shown rather moderate gains from the use of this local refinement technique, especially at low Reynolds numbers. The method suggested in [23] appears to be more efficient, although a direct comparison cannot be made since different test cases are treated. A factor which may limit the usefulness of local refinement is the zero-order accuracy of usual discretisation schemes at interfaces between different levels. This should be further investigated: The present tests suggest that this does not affect the overall 2$^{nd}$ order of accuracy with respect to refinement of the whole grid, but one suspects that the high truncation errors at level interfaces are indeed sources of additional discretisation error, thus limiting the gains from local refinement, especially at low Reynolds numbers because viscous terms contribute $O(1)$ to the truncation error while convection terms contribute $O(h)$. A possible way around this problem would be to construct more accurate discretisation schemes for the CVs which are close to the level interfaces. Such schemes would also be more expensive, but they would not result in significant overhead because the number of such CVs is very small compared to the total number of CVs of the domain (the dimensionality of the region covered by these CVs is one less than the dimensionality of the computational domain).

An important issue related to the efficiency of the local refinement technique is how to choose the local refinement parameter $r_\tau$ in (4.2). Normally $r_\tau$ should somehow be related to the required level of discretisation error, but this issue was not investigated in the present work. Alternatively, one may normalize the truncation error estimate by the main diagonal coefficients of the linearised discrete system as is done in [13], [24], to obtain a quantity which has the same dimensions as the discretisation error, and which is an indication of the contribution of each CV to the discretisation error. A difficulty with this approach is how to obtain an indication of the discretisation error of pressure since the continuity equation does not contain pressure terms.

The O(1) magnitude of the truncation error near level interfaces prevents the use of the local refinement criterion there, but in the present work this problem was overcome by simply not allowing local refinement near level interfaces. The same problem arises at boundary CVs, and there the problem is more serious because a suitable way must be found to identify regions where refinement is actually needed. Otherwise, the use of the local refinement criterion may result in the pile-up of local refinement levels near the boundaries, unnecessarily increasing the number of CVs and causing a large increase of the truncation error near the boundaries. A possible remedy is to use a refinement criterion which is based on the finite element residual (e.g. [23]) instead of the truncation error estimate. However, the truncation error estimate may also be used to obtain a more accurate solution (see [25]). An alternative remedy would be again to construct more accurate discretisation schemes for boundary CVs.

**TABLES**

| Grid | Re = 100 V(2,2)-1 | | Re = 1000 V(2,2)-1 | | Re = 1000 W(2,2)-1 | | Re = 5000 W(2,2)-1 | | Re = 10000 W(2,2)-1 | |
|---|---|---|---|---|---|---|---|---|---|---|
| 32×32 | 20 | *0.38* | 52 | *0.73* | 34 | *0.60* | 114 | *0.87* | 163 | *0.90* |
| 64×64 | 21 | *0.36* | 56 | *0.74* | 25 | *0.52* | 95 | *0.83* | 155 | *0.90* |
| 128×128 | 22 | *0.36* | 51 | *0.74* | 22 | *0.39* | 54 | *0.73* | 104 | *0.85* |
| 256×256 | 24 | *0.36* | 49 | *0.74* | 23 | *0.37* | 29 | *0.58* | 64 | *0.73* |

**Table 1:** Square lid-driven cavity, uniform grids: For each case, the left column shows the number of cycles of the classic SIMPLE/multigrid procedure ([8]) required to reduce the maximum residual per unit volume below $10^{-8}$, and the right column (in *italic*) shows the corresponding reduction factor (6.1). Full multigrid (FMG) is used in each case starting from the coarsest level 8×8.

| Grid | Re = 100 V(2,2)-1 | | Re = 1000 V(2,2)-1 | | Re = 1000 W(2,2)-1 | | Re = 5000 W(2,2)-1 $a_{mi}$=0.1 | | Re = 5000 W(2,2)-1 $a_{mi}$=0.2 CL=2 | |
|---|---|---|---|---|---|---|---|---|---|---|
| 32×32 | 20 | *0.35* | 56 | *0.76* | 36 | *0.60* | 130 | *0.88* | 226 | *0.93* |
| 64×64 | 20 | *0.36* | 53 | *0.75* | 25 | *0.49* | 124 | *0.87* | 156 | *0.90* |
| 128×128 | 22 | *0.37* | 52 | *0.75* | 22 | *0.39* | 127 | *0.88* | 91 | *0.84* |
| 256×256 | 23 | *0.37* | 48 | *0.78* | 22 | *0.36* | 134 | *0.88* | 73 | *0.77* |

**Table 2:** Like table 1, but using the modified SIMPLE/multigrid algorithm of Section 5.

| Grid | Re = 100 V(2,2)-1 | | Re = 1000 V(2,2)-1 | | Re = 1000 W(2,2)-1 | | Re = 5000 W(2,2)-1 | | Re = 10000 W(2,2)-1 $a_{mg}$=0.8 | |
|---|---|---|---|---|---|---|---|---|---|---|
| 32×32 | 20 | *0.35* | 57 | *0.75* | 36 | *0.62* | 157 | *0.90* | 235 | *0.94* |
| 64×64 | 20 | *0.36* | 55 | *0.76* | 25 | *0.49* | 114 | *0.86* | 219 | *0.93* |
| 128×128 | 22 | *0.37* | 54 | *0.76* | 22 | *0.39* | 51 | *0.73* | 176 | *0.92* |
| 256×256 | 23 | *0.37* | 49 | *0.78* | 22 | *0.36* | 29 | *0.54* | 119 | *0.88* |
| **CPU time** | 183 s | | 408 s | | 241 s | | 407 s | | 1519 s | |

**Table 3:** Like table 2, but with the modification of the pressure term of the local mass fluxes described in Section 6.1. Also shown is the CPU time for the whole FMG procedure on a 1.4 GHz Pentium 4.

| Grid | Re = 100 V(2,2)-1 | | Re = 1000 $W^k$/V(2,2)-1 | | | Re = 5000 $W^k$/V(2,2)-1 | | |
|---|---|---|---|---|---|---|---|---|
| **1 (5)** | 22 | *0.36* | 19 | *0.37* | $k=3$ | 113 | *0.86* | $k=3$ |
|  |  |  | 16 | *0.26* | $k=4$ | 62 | *0.74* | $k=4$ |
| **2 (6)** | 23 | *0.37* | 18 | *0.38* | $k=3$ | 112 | *0.86* | $k=3$ |
|  |  |  | 15 | *0.21* | $k=4$ | 64 | *0.76* | $k=4$ |
|  |  |  |  |  |  | 36 | *0.63* | $k=5$ |
| **3 (7)** |  |  | 17 | *0.38* | $k=3$ |  |  |  |
|  |  |  | 15 | *0.20* | $k=4$ |  |  |  |
| **4 (8)** |  |  | 16 | *0.36* | $k=3$ |  |  |  |
|  |  |  | 16 | *0.20* | $k=4$ |  |  |  |
| **CPU time** | 40 s | | 641 s | | $k=3$ | 399 s | | $k=3$ |
|  |  |  | 670 s | | $k=4$ | 298 s | | $k=4$ |
|  |  |  |  |  |  | 276 s | | $k=5$ |

**Table 4:** Square lid driven cavity, uniform grids: Convergence and CPU time for the FMG procedure on composite grids. The finest level of each composite grid is shown in parentheses, and the underlying grids are shown in Figures 8 and 9.

| Level | 3 (32²) | 4 | 5 | 6 | Total |
|---|---|---|---|---|---|
| CVs | 544 | 1816 | 352 | 256 | 2968 |
| Area | 33.55 % | 63.59 % | 2.83 % | 0.03 % | 100 % |

**Table 5:** Lid-driven cavity, non-uniform composite Grid 1: Number of global CVs of each level, and the area that they occupy.

| non-composite grid | | | | composite grid | | | |
|---|---|---|---|---|---|---|---|
| 32×32 | 47 | 0.71 | W(2,2)-1 | | | | |
| 64×64 | 34 | 0.61 | W(2,2)-1 | 1 | 49 | 0.74 | W²/V(2,2)-1 |
| 128×128 | 33 | 0.51 | W(2,2)-1 | 2 | 34 | 0.59 | W³/V(2,2)-1 |
| 256×256 | 46 | 0.70 | W(2,2)-1 | 3 | 47 | 0.71 | W³/V(2,2)-1 |

**Table 6:** Square lid-driven cavity, non-uniform grids, Re=1000: The table shows the numbers of cycles to drop the maximum residual per unit volume below $10^{-8}$ (left column), reduction factors (6.1) (middle column), and the type of cycle used in each case (right column).

| Grid | β = 45°, Re = 100 V(2,2)-1 | | β = 45°, Re = 1000 W(2,2)-1 | | β = 30°, Re = 100 V(3,3)-4 | | β = 30°, Re = 1000 W(2,2)-4 | |
|---|---|---|---|---|---|---|---|---|
| 32×32 | 17 | 0.29 | 52 | 0.70 | 12 | 0.23 | 20 | 0.39 |
| 64×64 | 18 | 0.29 | 38 | 0.64 | 15 | 0.28 | 22 | 0.46 |
| 128×128 | 19 | 0.29 | 23 | 0.47 | 16 | 0.37 | 15 | 0.33 |
| 256×256 | 20 | 0.29 | 19 | 0.30 | 22 | 0.49 | 14 | 0.26 |
| CPU time | 196 s | | 262 s | | 362 s | | 249 s | |

**Table 7:** Convergence of the skew lid-driven cavities problems on non-composite grids, and CPU time needed to complete the FMG procedure.

| Grid | β = 45°, Re = 100 V(2,2)-1 | | β = 45°, Re = 1000 W(2,2)-1 | | β = 30°, Re = 100 V(3,3)-4 | | β = 30°, Re = 1000 W(2,2)-4 | |
|---|---|---|---|---|---|---|---|---|
| composite 1 | 19 | 0.29 | 23 | 0.46 | 22 | 0.44 | 15 | 0.32 |
| composite 2 | 21 | 0.31 | 19 | 0.29 | 37 | 0.64 | 15 | 0.28 |
| CPU time | 91 s | | 178 s | | 286 s | | 156 s | |

**Table 8:** Convergence of the skew lid-driven cavities problems on composite grids, and CPU time needed to complete the FMG procedure.

| Grid | Re = 133 V(2,2)-1 | | Re = 133 V(2,2)-4 | | Re = 400 V(2,2)-1 | | Re = 400 V(2,2)-4 | | Re = 800 W(2,2)-4 | |
|---|---|---|---|---|---|---|---|---|---|---|
| 220×12 | 37 | 0.58 | 37 | 0.57 | 72 | 0.74 | 59 | 0.69 | 264 | 0.92 |
| 440×24 | 36 | 0.66 | 37 | 0.60 | 61 | 0.75 | 55 | 0.72 | 110 | 0.84 |
| 880×48 | 39 | 0.70 | 31 | 0.60 | 55 | 0.74 | 52 | 0.73 | 54 | 0.71 |
| 1760×96 | 55 | 0.86 | 40 | 0.69 | 85 | 0.89 | 48 | 0.73 | 48 | 0.73 |
| CPU time | 1087 s | | 1152 s | | 1642 s | | 1478 s | | 1992 s | |

**Table 9:** Convergence of the backward facing step problem on non-composite grids.

| Composite Grid | Re = 133 V(2,2)-4 | | | Re = 400 V(2,2)-4 | | | Re = 800 W(2,2)-4 | | |
|---|---|---|---|---|---|---|---|---|---|
| 1 | 33 | 0.56 | 26184 CVs | 52 | 0.73 | 30144 CVs | 57 | 0.72 | 36912 CVs |
| 2 | 36 | 0.61 | 61296 CVs | 50 | 0.75 | 69984 CVs | 42 | 0.72 | 93132 CVs |
| CPU time | 576 s | | | 952 s | | | 1579 s | | |

**Table 9:** Convergence of the backward facing step problem on composite grids.

**FIGURES**

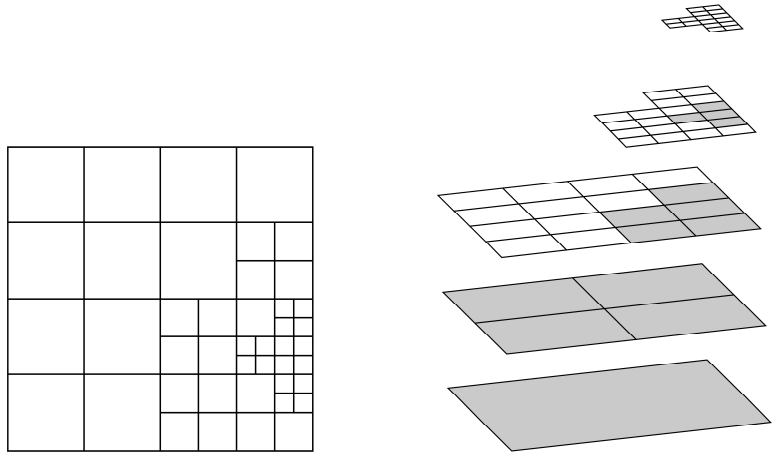

**Figure 1:** A composite grid and its analysis into levels. The local CVs are shown in grey.

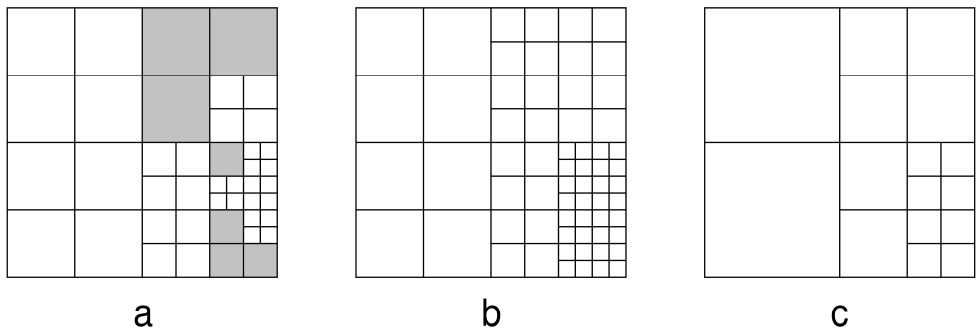

**Figure 2:** Grid (a) does not have an underlying grid because the grey CVs are global siblings of local CVs. Their refinement results in grid (b) whose underlying grid is (c).

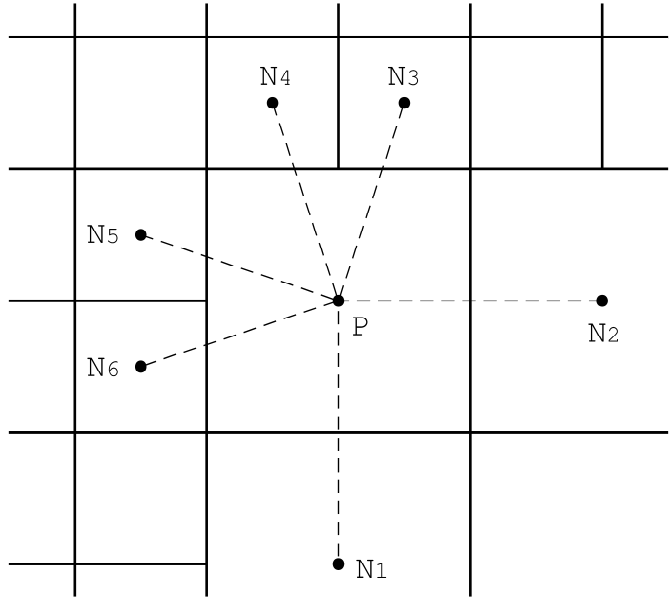

**Figure 3:** A CV *P* with exterior faces and its neighbours.

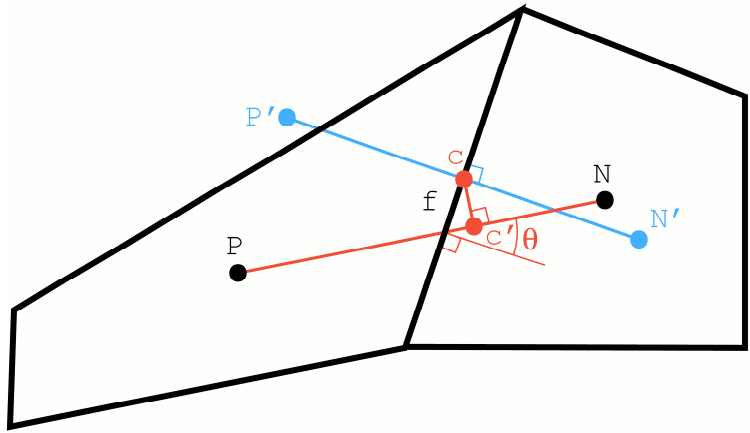

**Figure 4:** A face *f* separating CVs *P* and *N*, and related notation.

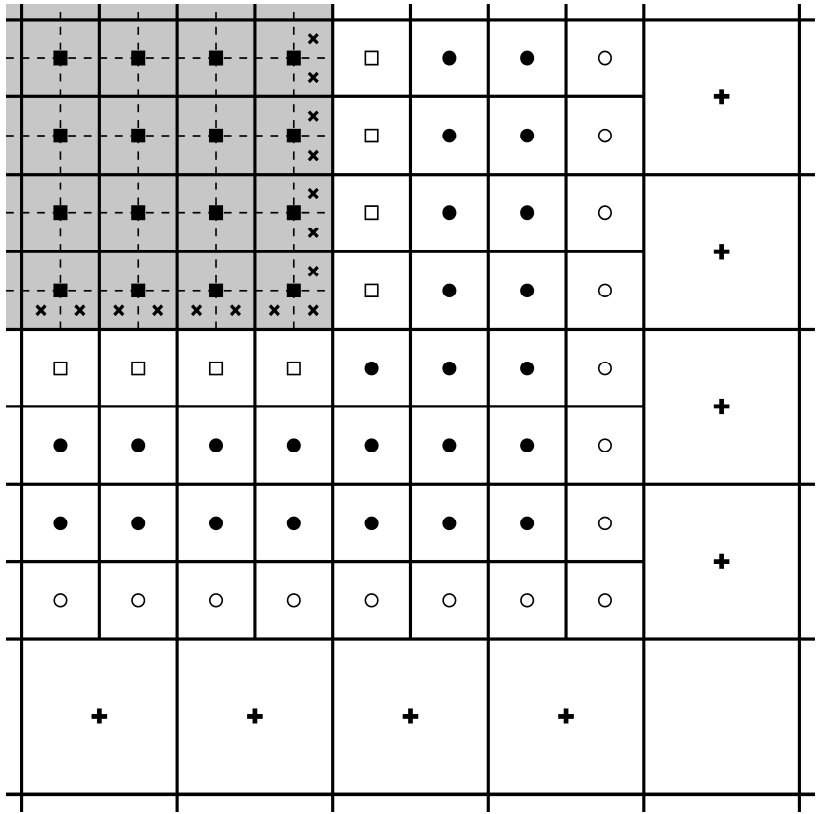

**Figure 5:** Part of a composite grid displaying 3 levels, with emphasis on the middle level. The local part of the middle level is shown in grey and the finest level is shown in dashed line. The centres of the CVs of the middle level are also marked, as are the centres of the neighbouring CVs of the other levels.

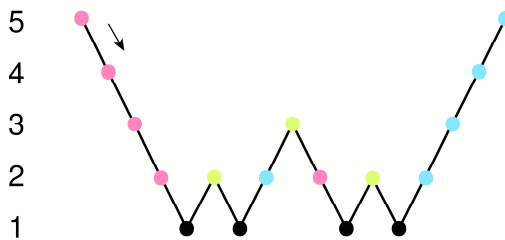

**Figure 6:** Schematic representation of the $W^3/V(v_1,v_2)$ cycle on a grid with 5 levels. At red, blue, green dots $v_1$, $v_2$, $v_1+v_2$ SIMPLE relaxations are performed respectively. At black dots either direct solution or many relaxations are performed.

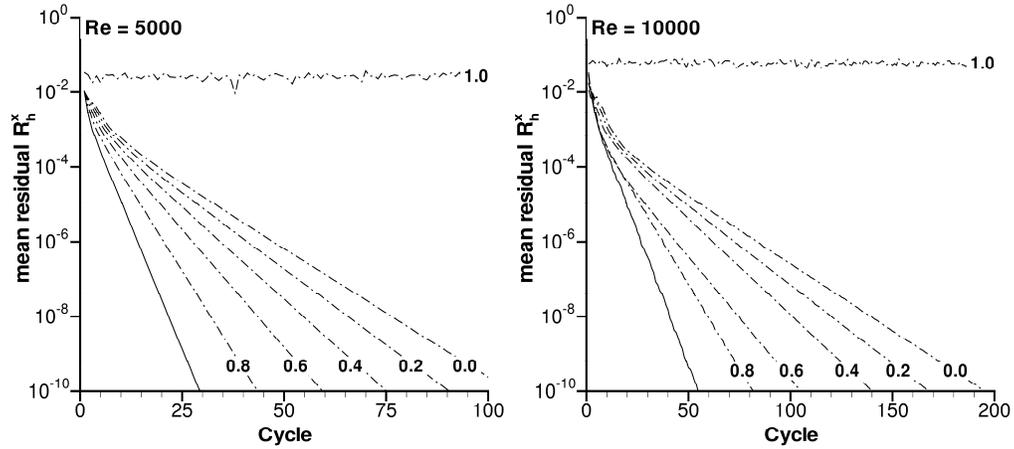

**Figure 7:** Square lid driven cavity, 256×256 grid: Convergence histories of the *x*-momentum residuals per unit volume using the classic SIMPLE/multigrid method, without FMG. The solid line corresponds to smoothing of corrections according to (5.15), and the chained lines to the use of CDS/UDS blending for the local momentum fluxes with the blending factor $a_c$ indicated on each curve, without correction smoothing.

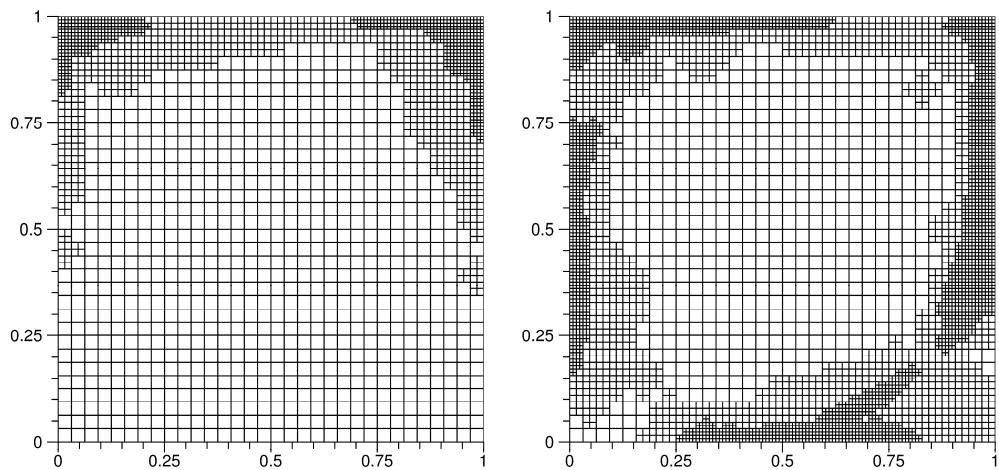

**Figure 8:** The underlying grids of the final grids for Re=100 (left) and Re=5000 (right).

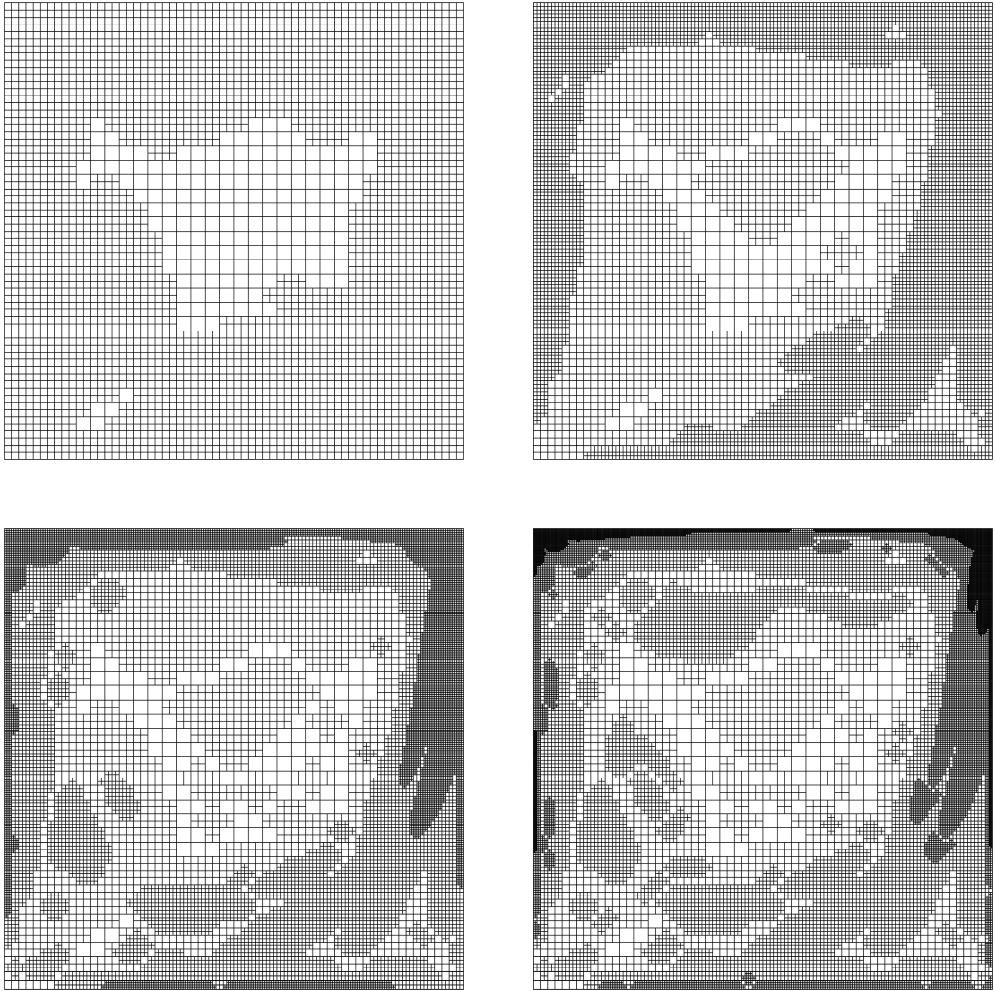

**Figure 9:** The underlying grids of grid 1 (top left), grid 2 (top right), grid 3 (bottom left) and grid 4 (bottom right) for Re = 1000.

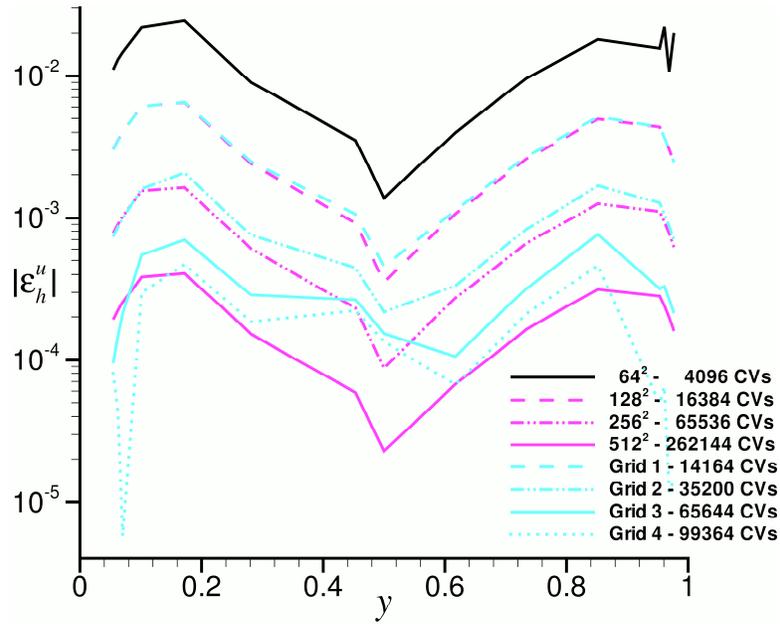

**Figure 10:** Absolute value of $\varepsilon_h^u$ along the vertical centreline of the cavity, for various non-composite (purple lines) and composite (cyan lines) grids, Re = 1000.

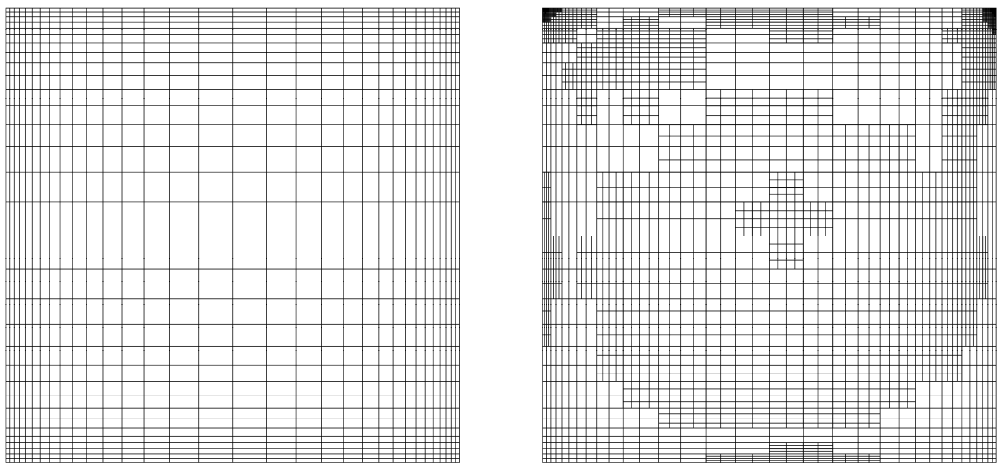

**Figure 11:** The 32×32 non-uniform grid (left) and non-uniform composite Grid 1 (right).

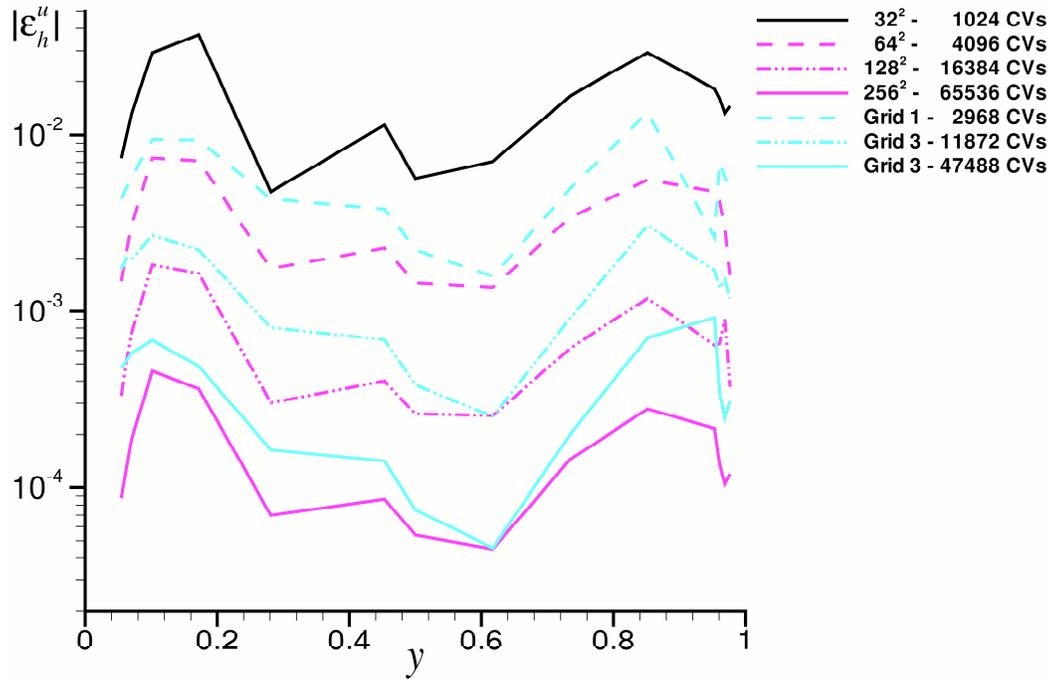

**Figure 12:** Absolute value of $\varepsilon_h^u$ along the vertical centreline of the cavity, for various non-composite (purple lines) and composite (cyan lines) non-uniform grids, Re = 1000.

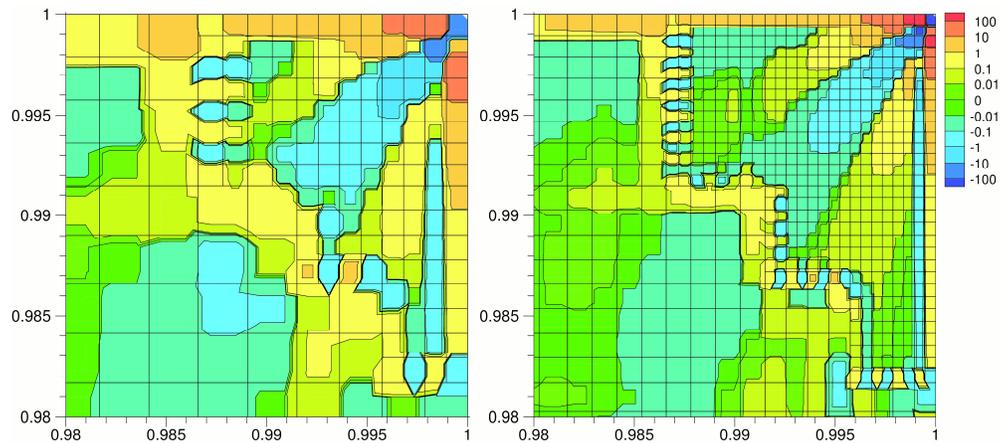

**Figure 13:** Estimate of $\tau_h^x$ on part of the non-uniform composite grids 2 (left) and 3 (right), Re = 1000. The underlying grids are also shown.

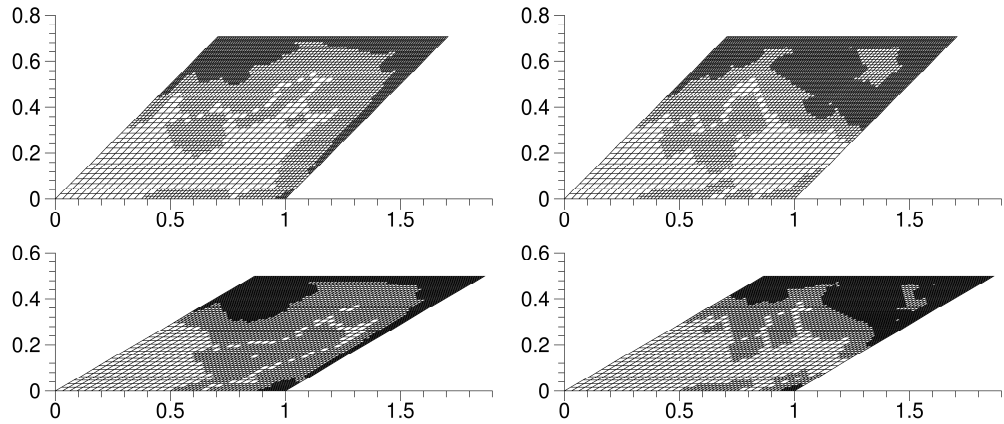

**Figure 14:** The underlying grids of the final grids for Re = 100 (left), Re = 1000 (right). Top: $\beta = 45^\circ$; Bottom: $\beta = 30^\circ$.

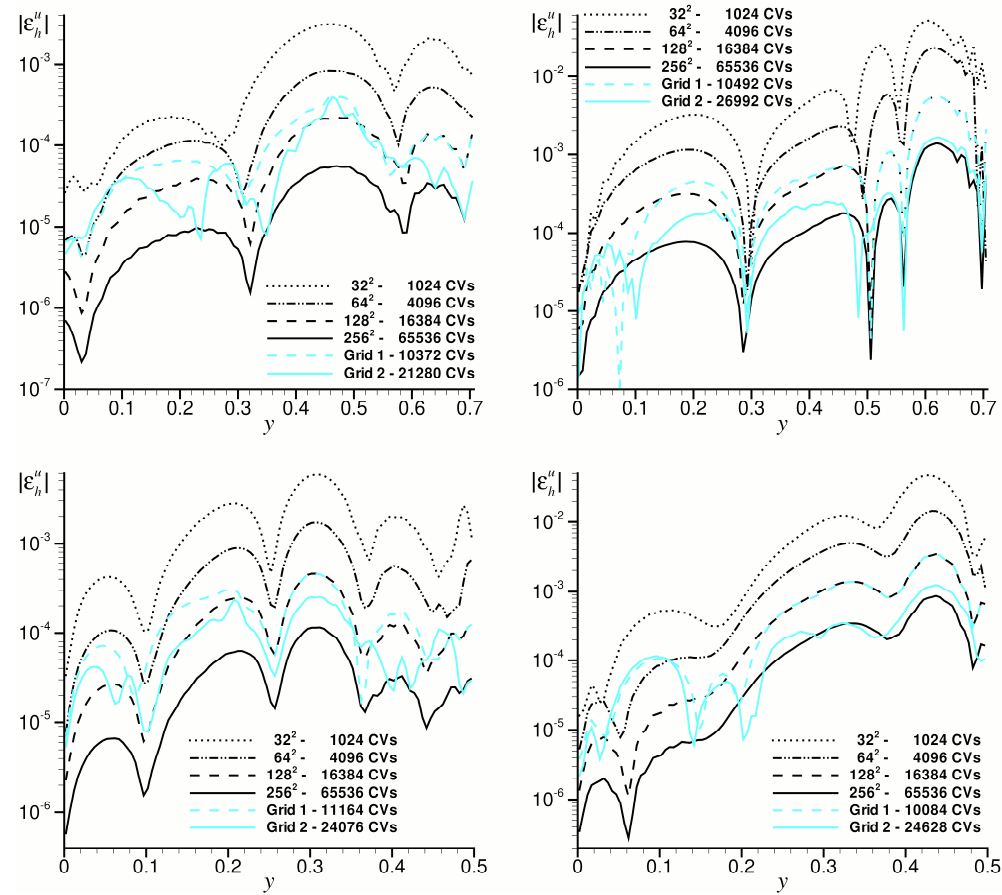

**Figure 15:** Absolute value of $\varepsilon_h^u$ along the centreline of the skew cavities which is parallel to the side walls. Top: $\beta = 45^\circ$; Bottom: $\beta = 30^\circ$. Left: Re = 100; Right: Re = 1000.

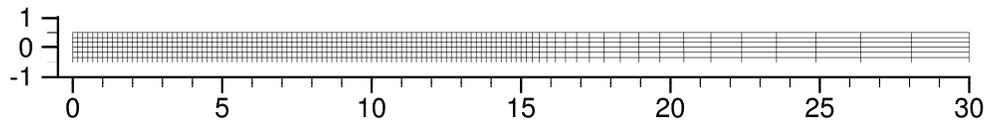

**Figure 16:** Level 1 for the backward facing step problem.